# *Towards the Adoption of OMG Standards in the Development of SOA-Based IoT Systems*

Bruno Costa[1], Paulo F. Pires[2], Flávia C. Delicato[2]
[1]Instituto Federal de Educação, Ciência e Tecnologia do Rio de Janeiro (IFRJ), Rio de Janeiro, Brazil
[2] Instituto de Computação (IC), Universidade Federal Fluminense (UFF), Rio de Janeiro, Brazil
{brunocosta.dsn, paulo.f.pires, fdelicato}@gmail.com

*Abstract*— **A common feature of the Internet of Things (IoT) is the high heterogeneity, regarding network protocols, data formats, hardware and software platforms. Aiming to deal with such a degree of heterogeneity, several frameworks have applied the Model-Driven Development (MDD) to build IoT applications. On the software architecture viewpoint, the literature has shown that the Service-Oriented Architecture (SOA) is a promising style to address the interoperability of entities composing these solutions. Some features of IoT make it challenging to analyze the impact of design decisions on the SOA-based IoT applications behavior. Thus, it is a key requirement to simulate the model to verify whether the system performs as expected before its implementation. Although the literature has identified that the SOA style is suitable for addressing the interoperability, existing modelling languages do not consider SOA elements as first-class citizens when designing IoT applications. Furthermore, although existing MDD frameworks provide modeling languages comprising well-defined syntax, they lack execution semantics, thus, are not suitable for model execution and analysis. This work aims at addressing these issues by introducing IoTDraw. The framework provides a fully OMG-compliant executable modeling language for SOA-based IoT systems; thus, its specifications can be implemented by any tool implementing OMG standards.**

## 1 Introduction

Heterogeneity is an intrinsic feature in the Internet of Things (IoT) systems, regarding several aspects such as network protocols, data formats, hardware and software platforms [1]. Such a degree of heterogeneity poses challenges to develop IoT applications when following traditional approaches, as the node-centric programming [2], which has been widely used in the context of Wireless Sensor Networks (WSN) [3], a subset of IoT. In such an approach, applications are developed in a platform-dependent way, with experts in embedded systems programming the tasks required for individual nodes, as well as the interactions between them, by using general-purpose programming languages. As shown by [4], although traditional approaches allow complete and fine-grained control over individual nodes, it is not easy to apply them in the context of IoT applications due to the scale factor and high heterogeneity of devices [5]. In addition to these factors, programming using traditional approaches requires low-level knowledge of the hardware-specific operating systems of the sensor nodes. The proliferation of sensors and applications that make use of their data has led to the desire to increase the learning curve for building sensing-based systems and to promote the reuse of generated software artifacts. Aiming at addressing the limitations of traditional approaches, several frameworks were proposed (such as [6], [7]) applying the Model-Driven Development (MDD) [8] to build IoT and large-scale WSN applications.

MDD is a development paradigm that uses models as the primary artifact of the development process. With MDD approaches, models specifying the systems are created by using Domain-Specific Modeling Languages (DSML), or DSL for short. A DSL provides high-level abstractions related to the domain of interest expressed through inter-connected textual or graphical symbols. In the context of IoT, existing DSLs allow identifying the heterogeneous entities composing the applications through general concepts. As suggested by the literature (e.g., [6], [9], [10]), considering its key principles, such as abstraction, separation of concerns, reusability and automation, MDD is a promising paradigm to support the development of IoT solutions.

From the software architecture point of view, the heterogeneous nature of IoT requires structuring the applications by considering the interoperability of different software components and hardware devices. This critical issue has led the researchers to investigate architectural patterns or styles that most favor the communication of the inter-connected entities in an IoT system. In this sense, many studies (e.g., [11], [12]) have shown that the Service-Oriented Architecture (SOA) [13] is one of the most suitable architectural styles to address the interoperability in IoT applications. Interoperability is about the degree to which two or more systems can usefully exchange meaningful information via interfaces in a given context [14], [15]. SOA is an architectural pattern structured as a collection of loosely coupled, distributed components that provide and/or consume services through well-defined interfaces [15]. In the SOA-based IoT applications, the devices' capabilities (e.g., sensing, actuating) are provided as software services through such well-defined interfaces, thus allowing the entities that compose an IoT providing their capabilities and consuming the required services in a standardized manner. For example, the authors in [16] applied SOA in real-world scenarios with multiple heterogeneous devices, ensuring interoperability across all devices and components. Furthermore, Cloud platforms are widely used as back-end for IoT given their huge processing and storage capabilities. The synergistic integration of IoT with Cloud has given rise to the recent paradigm known as Sensor-Cloud or Cloud of Things, which aims to make the most of both technologies. IoT-generated data that requires a lot of processing or long-term storage can be sent to the cloud, which in turn extends its service portfolio to encompass sensing and actuation capabilities. Cloud platforms follow a service-based resource provisioning model. Cloud platforms providing infrastructure for IoT systems naturally adopt the services approach (e.g., [17], [18]). In a recent survey that investigated middleware technologies for Cloud of Things [19], the authors have identified that SOA is the main architectural style in CoT-based Middleware, due to is capability of promoting interoperability and reusability in IoT systems.

### 1.1 Issues in designing MDD & SOA-based IoT Systems

Any DSL must provide a coherent set of concepts related to the domain of interest [8]. That is, the DSL must be expressive, aiming to allow the precise identification of multiple aspects regarding the application domain. Otherwise, if the DSL lacks essential elements of the underlying domain, the models may be at least ambiguous, or the modelers may not be able to represent all components of the applications. Although the literature has identified the suitability of the SOA style for addressing the interoperability in the IoT domain, to the best of our knowledge, there is no DSL tailored for the modeling of SOA-based IoT applications. SOA is a proven architectural pattern, with a well-known set of concepts. Existing DSLs for IoT (e.g., UML4IoT [9] and ThingML [20]) do not fully contemplate such concepts, which impacts on the architectural design of IoT applications following the SOA approach. This is a critical problem since even crucial but straightforward design questions related to SOA style cannot be answered by the model, such as, "What are the services, and by which interface they are exposed?", "Who are the provider/consumer participants that are part of the application?", "How the participants interact with each other?", "What is the choreography of the services?", or "What is the protocol used by the interfaces." A DSL that does not provide concepts to allow answering these design questions possibly hampers the design of SOA-based IoT applications.

Another important design issue refers to the simulation and analysis of IoT systems. In the domain of Software and Systems engineering, a rule of thumb states that the best system representation is the simplest model that answers a set of design questions [21]. This means that, besides identifying precisely the multiple components, another reason for modeling a system is to support the architectural decision-making process [21], [22]. Otherwise, the system model would serve only as documentation and would not support a more in-depth analysis, which may not justify the modeling effort. A modeling approach helps to answer design questions and supports the architectural decision-making process by its ability to predict, at design-time, the properties of an artifact based on its design. This prediction is possible when the DSL allows its compliant models to be executed and simulated. When specifying IoT systems, some features of the IoT domain makes challenging to analyze the impact of design decisions on the system behavior. Examples of such features are the resource constraints of devices and multiple deployment scenarios. An IoT application can be composed of tens, hundreds, or even thousands of devices, many of them being powered by limited batteries and with constrained resources (in terms of processing speed and memory size), which may impact on the availability or performance of the IoT application [1]. On the other hand, the multiple components of an application may be deployed on different platforms (varying from resource-constrained devices to Cloud computing platforms), which result in a considerable number of eligible deployment scenarios. Since each deployment scenario has a different impact on application requirements and system performance, it may become humanly infeasible to identify the best deployment alternative. These issues hamper making proper design decisions, being necessary to execute and simulate the model to verify whether the specified application performs as expected before its implementation. Otherwise, if problems caused by wrong design decisions are found only after the system implementation or deployment, it may result in financial loss or, worse, risks to human lives, for instance in failures in e-Health applications.

Considering the system verification and the problems that may occur whether it is not realized at design time, a typical analysis that must be performed refers to non-functional or Quality of Service (QoS) properties. The study of Udoh and Kotonya [23] presents a comprehensive review of existing IoT application frameworks and toolkits. An interesting finding of such study is that existent frameworks ignore the support of verification of Quality of Service (QoS) properties at design time. Another study ([24]) has shown that only around 2% of the companies that are developing IoT solutions are able to test the specified systems before undertaking the system deployment.

### 1.2 Contributions and Roadmap

Aiming to tackle the abovementioned issues, the core contribution of our work is proposing an MDD framework, which we called as *IoTDraw*. The framework provides a fully OMG-compliant executable modeling language (DSL), *SoaML4IoT* [25], for SOA-based IoT systems; thus, its specifications can be implemented by any tool implementing OMG standards. The DSL was built on the Service-oriented architecture Modeling Language (SoaML) [26], the Object Management Group (OMG) standard for designing SOA solutions. SoaML consists of an extension of UML [27] (i.e., a UML profile), which was conceived through the collaboration of experts from academia and industry. As stated by the SoaML specification, the modeling language focuses on the basic service modeling concepts, and the intention is to use it as a foundation for further domain extensions. Thus, based on our experience and proven domain models for IoT (e.g., IoT-ARM [11], WSO2 [28]), we extended this language with concepts of the IoT domain, allowing the identification of the multiple entities composing SOA-based IoT applications.

SoaML4IoT-compliant models are meant to be executed based on the language execution semantics. The execution semantics of SoaML4IoT is defined following the fUML (Foundational Subset for Executable UML Models) [29], the OMG standard for model execution. This specification provides a semantic model that formalizes the execution of a subset of UML, that is, classes (structure) and activities (behavior). Therefore, fUML allows defining an interpreter or execution engine to execute the system described in the model. However, if the UML model has an applied profile (e.g., SoaML), the semantics of this latter does not influence the execution. Aiming to allow the execution of SoaML4IoT models, we extended the fUML semantic model with concepts of our proposed DSL and implemented it as a prototype on Moka [30], a Papyrus module that includes an execution engine complying with fUML.

By allowing the execution of SoaML4IoT-compliant models, our proposed framework aims to help answering design decisions that are hard to answer even for simple IoT applications. Some of these questions are, for example: "In which platform should I deploy this component?" or "Considering the availability requirement of the application, is there any component I should deploy to a different platform after a link failure?". By supporting answering these questions, IoTDraw aids modelers to make more successful architectural decisions and to verify their impact on the IoT application behavior/performance.

We claim that our proposal is useful for supporting engineering activities potentially for all IoT application domains, including the field of Industrial Internet of Things (IIoT) systems [31]. As identified by Koziolek and colleagues [32], distributed control systems are currently evolving towards IIoT. However, the initial network configuration, application-specific device configuration, and integration with other devices are cumbersome and laborious nowadays. Such tasks are referred to as *commissioning*, and engineers still suffer from complex commissioning process that incurs high costs, not only from investments on hardware components but, also from manual labor to engineer and set them up [33]. The number of heterogeneous devices (hundreds or even thousands) composing the system, the specificity of communication protocols, as well as the eventual reconfiguration of devices can result that the commissioning process takes several months, which is costly and delays time-to-market. The most common structure of control systems follows the ANSI/ISA95 [34] standard, which splits the systems into four levels, namely, Level 4: Business Planning & Logistics (plant production scheduling, operational management); Level 3: Manufacturing Operations & Control (dispatching production, detailed production scheduling, reliability assurance); Level 2: Supervision and Controller (workstations for supervision by human operators and for engineering of devices and processes); and Level 1: Devices (sensors and actuators that directly interface with the machines executing the industrial process). The IoTDraw can be used in the commissioning process, specifically, in the levels 1 to 3 of the ANSI/ISA95 standard. In Level 1, the SoaML4IoT allows the engineers to specify the configuration of devices through UML elements thus, abstracting hardware-specific details. Such device models can be organized into model libraries and re-used by other projects. In Level 2, the controllers and their network connection to the devices are also modeled by using UML. By applying SOA standard, the distributed components are structured as providers and consumers, which expose their functionalities through well-defined interfaces, promoting interoperability. Finally, the operations of Level 3 are specified as behavior UML diagrams (i.e. UML Activity Diagram) in such a way that it is possible to execute the model at design-time. Thus, engineers can verify and test the system before its implementation and deployment. Such verification can reduce the cost of fixing errors that would be only detected in the production.

IoT applications interact with the physical world through different models, such as, *periodic*, *event-driven*, or initiated-by-the-actor (i.e., *request-response*) [1], [35]. In the periodic model, the devices sense data (e.g., temperature, lighting) or actuate on the physical objects (e.g., turning on/off, updating configurations) continuously, at a rate predefined by the application. In the event-driven model, the devices continuously monitor the physical objects or environment variables, but report information or perform actuation only if an event of interest for the application occurs. The event-driven model requires asynchronous communication, for example, based on the Publish-Subscribe pattern [15]. With such a pattern, the applications register for events of interest only once and receive the required sensing data upon the occurrence of these events. Finally, in the request-response model, the applications send requests for sensing data or actuating capabilities to the devices, which perform the required tasks and respond to the applications with the required data or actuating confirmation. The request-response model is based on synchronous communication, in which the devices report their data in response to a synchronous request issued by the application. The focus of IoTDraw is on periodic applications, which, as indicated by [36], is adopted in several domains and IoT solutions.

In summary, the contributions of our work aim to benefit both the IoT and MDD fields. In the IoT domain, we bridge the gap of existing DSLs that do not provide concepts as first-class citizens for the precise representation of periodic IoT systems following the services approach and thus, lack support for specifying and simulating SOA-based periodic IoT applications. Considering the MDD field, our work is one of the first endeavors towards defining the execution semantics for UML profiles by extending the fUML semantic model. Since IoTDraw is fully OMG-compliant, its specifications can be implemented by any tool implementing UML and fUML.

The remainder of this paper is organized as follows. Section 2 establishes the background underlying this work. Section 3 describes a Smart City IoT system that will be used as a running example of this study. Section 4 introduces the IoTDraw modeling framework. In Section 5 we present the evaluation of our proposal. In Section 6 we analyze the related work. Finally, Section 7 revisits the contributions and presents perspectives for future work.

## 2 Foundations

In this Section, we present an overview of the basics of MDD engineering, with focus on the elements on which our approach is built on, that is, SoaML and fUML.

### 2.1 SoaML

SoaML [26] is the standard graphical modeling language proposed by the OMG for designing systems that follow the services approach. It consists of a metamodel and a UML profile for the specification of services within SOA style. An extensive example of modeling with SoaML is given by Elvesæter and colleagues [37]. In the following, we present key concepts of the SoaML metamodel. In parenthesis, we show the UML element that SoaML profile extends.

*Participants* (Class) are entities that provide or use services. *Capability* (Class), refer to functions required by stakeholders and provided by a participant through a service. A *Service* (Port) denotes value delivered to another through a well-defined interface. *Provider* and *Consumer* (Interface) are roles of participants that provide or consumes services, respectively. The description of how the participants interact to provide or use a service can be specified as *Service Contracts* (Collaboration). It specifies the roles each participant plays in the service – provider or consumer – and the choreography of the service, that is, what information is sent between the provider and consumer and in what order. When specifying the choreography of a given service contract, any UML behavior specification can be used, such as interaction and activity diagrams. Finally, the *Services Architecture* (Collaboration) describes how participants work together for a purpose by providing and using services expressed in the service contracts.

There are a variety of approaches for identifying services that are supported by SoaML. However, regardless of how the services are identified, they are formalized by service descriptions. In turn, the services architecture aims at structuring such services with the interacting participants and their ports, which realize the roles of the system.

#### 2.1.1 SoaML and Microservices

Microservices is an architectural style that has gaining popularity in the last years. Emerging out of Service-Oriented Architecture [38], microservices is an approach that inherits many features of SOA [39], such as encapsulation (separation of implementation from interfaces, this last providing the service's functionalities), loose-coupling (minimizing dependencies between system components), and interoperability (communication through well-defined interfaces). Complementary to such features, the microservices style introduces novel concepts that differ it from SOA. Among them, the literature [40] highlights the *service granularity*, on which the microservices are more fine-grained than SOA services, with a single responsibility and owning its own data store. Such a feature allows the microservices being deployed independently and being replaceable in a smoother fashion than SOA components.

Considering the architectural design, microservices also differs from the SOA style. SOA-based systems are commonly decomposed into horizontal technical layers (a.k.a., horizontal decomposition), on which the services are grouped according to their functional similarity (e.g., data services, business services, or presentation services). On the other hand, in microservices architecture (MSA) each microservice should align to a single business capability, encapsulating all relevant technical layers [40], i.e., the vertical decomposition.

The SoaML focuses on the basic service modeling concepts, and, as stated in the language standard [26]), the goal is to use it as a foundation for further domain extensions. In this sense, SoaML can be extended to provide concepts of microservices (e.g., granularity, independent deployment). For example, [41] introduces major challenges of microservice design and presents ways to cope with them based on model-driven development. The authors propose an UML profile that extends SoaML elements to allow the specification of microservices-based systems. As an example, the concept of "Microservice" is modeled as a stereotype with the same name that specializes the SoaML Participant element.

### 2.2 Execution Semantics and fUML

The execution semantics of a DSL can be defined implicitly or explicitly [42]. Considering the implicit specification, a commonly used approach to making the conforming models executable is to develop model-to-code translators. Such translators map the DSL's metamodel concepts into coding elements of the general-purpose programming languages and generate the code from the DSL's conforming models. In this approach, the behavioral semantics is defined implicitly, encoded in the manually developed translators. The implicit specification has considerable drawbacks, as extensively demonstrated by the literature [42]. For example, the behavioral semantics might be redundantly defined, as it is necessary to create multiple model-to-code translators to various target languages. To overcome the implicit specification limitations, the behavior semantics of DSLs must be defined explicitly. In this specification, the DSL's execution semantics is specified by determining the steps of computation required for executing a conforming model. Therefore, the explicit specification expresses runtime concepts, which define an interpreter for the DSL. As demonstrated by Tatibouët and colleagues [43], fUML can be used to formalize the execution semantics of UML profiles explicitly. The execution semantics of fUML is defined by the fUML semantic model, which, in turn, is defined using the PSL [44]. The execution model is based on the Visitor pattern [45], which is used to add behavior to an already existing class hierarchy. In the case of fUML, the visitor pattern is applied over the UML subset (i.e., classes and activities), providing a specification for the execution of models represented in terms of instances of this subset. Formalizing the execution semantics of UML profiles consists of extending such visitors for each concept of the profile by using standard object-oriented mechanisms (e.g., inheritance, polymorphism).

## 3 Running Example

As adopted by several players (e.g., Google [46]) and identified by the literature (e.g., [47]), a common end-to-end infrastructure for IoT encompasses typically three tiers, namely, *cloud*, *fog*, and *device*. The cloud tier consists of computers/data centers located in the cloud providing resources (e.g., computing, storage), which can be rapidly provisioned and released on-demand. The fog tier encompasses computers that exploit capabilities at the edge of the Internet, thus, providing near-devices computing/storage resources [48]. Fog nodes can also act as bridges to connect the devices to the cloud and provide remote access to devices through APIs offered by specific components [1]. It is important to emphasize that in many implementations, the Fog tier assumes a hierarchical, multi-level organization. Therefore, the resulting system actually has multiple tiers in terms of possible component deployments. In turn, the device tier refers to electronic platforms equipped with sensors and actuators that are attached to physical entities to enhance them with sensing, actuating, communication, and computing capabilities.

Our running example is based on a real Smart City IoT project deployed in the city of Padova (Italy) that adopts the infrastructure presented above. The "Padova Smarty City" (PSC) [49] is the result of the collaboration between the municipality of Padova, the University of Padova, and the Patavina Tech – a software house specialized in the development of innovative IoT solutions. The infrastructure of PSC is depicted in Figure 1 and explained in the following. The complete explanation, including the reasoning behind all technical decisions, can be found in [49].

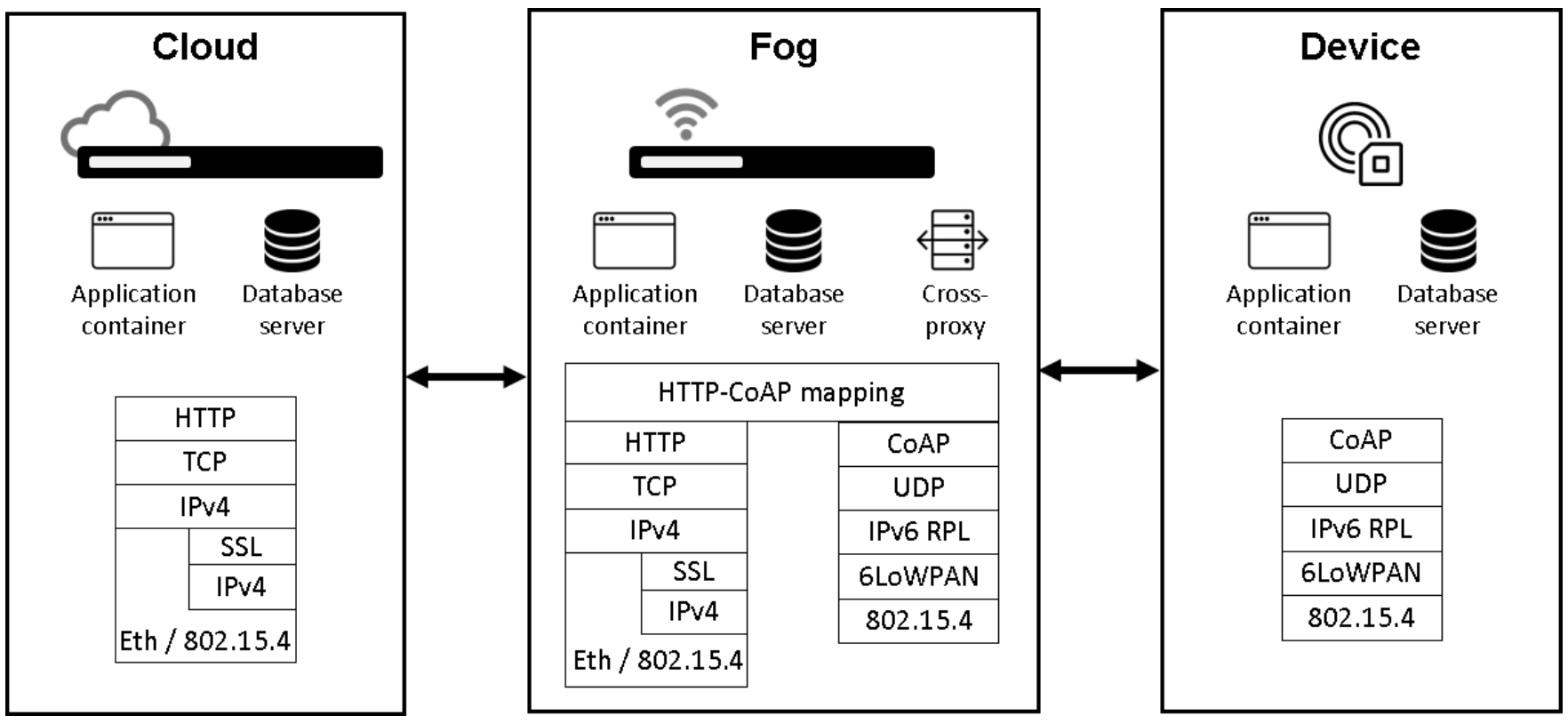

Figure 1. The infrastructure of "Padova Smart City" (based on [49])

The PSC encompasses several devices placed on streetlight poles of Padova's downtown and connected to the municipality's network through fog nodes. Some devices are heterogeneous, i.e., equipped with a photometer, temperature sensor, and humidity sensor. There are two types of homogenous devices. One is equipped only with benzene ($C_6H_6$) sensor, which monitors the air quality. The other is equipped with water sensor and placed on the base of the streetlight poles aiming to monitor the water level is case of flooding. A small battery powers IoT devices equipped with a photometer, temperature sensor, humidity sensor, and water sensor while the devices equipped with benzene sensor, which requires a much higher power supply, are powered by the grid. The devices have a CC2420 transceiver that implements the IEEE 802.15.4 standard [50], providing them with wireless communication. Cloud nodes refers to third-party computers that provide resources that can be provisioned or released on demand.

The devices, fog and cloud nodes are computers that provide database servers and application containers, allowing data storage and execution environment for software components, although, in the cloud, such capabilities can be provisioned and released on-demand. The fog nodes also provide protocol translation and functional mapping between unconstrained protocols and their constrained counterparts, which are used by the devices (e.g., CoAP [51], 6LoWPAN [52]), through a *cross-proxy*. A fog node performs protocol translation and forwards the requests directly to the required device by using 6LoWPAN and the RPL routing protocol [53]. Such a protocol is used in the communication between fog nodes and devices.

### 3.1 IoT Applications

On the PSC infrastructure, we want to design IoT applications to enhance the quality of the services offered to citizens. The applications are Flooding Warning (FW), Street Light Monitoring (LM), Temperature Monitoring (TM), Humidity Monitoring (HM), and Air Quality Monitoring (AM). The goal of these applications is to request the related data at regular time intervals (i.e., *periodic applications*), store it in a database, and further perform analytics aiming at examining different scenarios, which will be helpful for urban planning.

The SOA style is suitable for architecting the applications for two main reasons. Firstly, it allows providing the capabilities of the heterogeneous devices composing the applications through well-defined interfaces, thus, furnishing the interoperability between the different hardware platforms and software components. Secondly, some sensors will also be exposed as a service to be requested by external applications based on the *request-response* model, following the Sensing as a Service ($S^2$aaS) approach [54]. In $S^2$aaS, the sensing data is available for external users and can be provided on demand based on a *pay as you go* model, in the same way as traditional services. In short, when a user needs data of a certain object or environment, he/she requests it to the device's service, which answers with the required data. For example, in the context of the FW application, by providing water level sensor as a service, public/private organizations or individual developers can create their own flooding monitoring applications. For example, the University of Padova may create an application aiming to measure the time it takes for an area to be flooded after a storm starts. On the other hand, Patavina Tech may create a mobile app that informs the citizens about flooded areas.

Constrained batteries power the devices providing the lighting, temperature, humidity, and water level data; thus, a Quality of Service (QoS) requirement that must be analyzed in the related applications is the operational *lifetime* of devices providing the required services. Based on [55], we consider the lifetime as the time spanning from the instant when the device starts functioning until it runs out of energy, making the service provided by such a device unavailable. In this sense, the ability of the service, platform, or component to perform its required function over an agreed period may be impacted. Thus, the FW, LM, TM, and HM applications shall address *availability* requirements [15].

Flood monitoring is a time-critical application; thus, the *response time* requirement must also be addressed by the FW application. Based on [56], we consider the response time as the time spent since a component requests for a device's service until this component receives the sensing data or the device performs the required actuation.

Finally, considering the $S^2$aaS model, there is also a critical concern for the service consumers, namely, the *freshness* of the data. Data freshness refers to the time elapsed since the data is collected until it is delivered to the requesting user [57]. Some applications may require data as fresher as possible. This is the case, for example, of mobile apps that inform the citizens about flooded areas. On the other hand, other applications do not need a high data freshness, for example, applications that request the water level to perform historical data analysis.

A DSL providing concepts of the SOA style is required for the precise specification of the applications, with their underlying infrastructure. Moreover, another feature that a modeling language must provide is the support for model execution without requiring translating the model to other languages, avoiding the problems of translational approaches. Such a feature of the language should help answering design questions and support the architecture decision-making process. For example, a crucial decision is "in which cloud or edge nodes each component should be deployed?" To answer such a question, the modelers may evaluate dozens or even hundreds of candidate deployment scenarios. This is because more than one component would be able to be deployed within the same platform. The deployment decision becomes even more complicated considering that depending on the platform in which the components are deployed; it may impact on the availability or response time requirements. In this context, selecting the best candidate deployment scenario becomes humanly infeasible as the number of eligible deployments changes.

## 4 IoTDraw Modeling Framework

In this section, we present our proposed modeling framework, the IoTDraw. As introduced earlier, IoT Draw is more suitable for the specification and analysis of periodic applications. The actual version covers essential concepts or services-oriented architecture; however, since our approach was conceived on SoaML, it can be extended to allow the representation of microservices elements in the context of IoT systems.

SoaML4IoT is composed of two main cornerstones, namely, the SoaML4IoT modeling language (encompassing the abstract syntax, the concrete syntax, the execution semantics, and the extensibility rules) and the Model Execution Engine. The modeling language was designed by following the methodology presented by Brambilla and colleagues [8], which introduces five principles that a DSL should follow in order to be useful, that is, (i) *the language must provide good abstractions to the developer, must be intuitive, and make life easier, not harder*; (ii) *the language must not depend on one-man expertise for its adoption and usage*; (iii) *the language must evolve and must be kept updated based on the user and context needs*; (iv) *the language must come together with supporting tools and methods*; and, (v) *the language should be open for extensions and closed for modifications*. As introduced in the next subsections, SoaML4IoT was conceived considering these principles. The Execution Semantics and the Model Execution Engine were developed based on the process of formalizing the execution semantics of UML profiles proposed by Tatibouët and colleagues [43], which was briefly introduced in Section 2.2.

### 4.1 Abstract Syntax

The SoaML4IoT metamodel is depicted in the UML Class diagram of Figure 2. It was conceived as an extension of the SoaML metamodel (gray elements). Following the first principle of DSL's, the concepts related to the IoT domain were elicited from proven domain models for IoT, namely, the IoT Architectural Reference Model (IoT-ARM) [11], the WSO2 IoT Reference Architecture [28], and the IEEE Standard for an Architectural Framework for Internet of Things (P2413) [58]. Such domain models provide high-level abstractions that aim at describing the main concepts of IoT.

An *IoT System* is a cyber-physical system composed of *Platforms*, *Applications*, and *Networks*. A platform refers to computer nodes, which can be of type *Cloud*, *Fog*, or *Device*. Nodes located in the cloud provide a set of resources (i.e., processing, storage), which can be provisioned and paid on demand. Fog nodes are computers that act as near- devices computing/storage resources or as bridges to connect the devices to the cloud. A device is a computer attached to *physical entities*, which can be anything of the real world from objects and cars to animals and human beings. A physical entity lies in a specific geographic *location*.

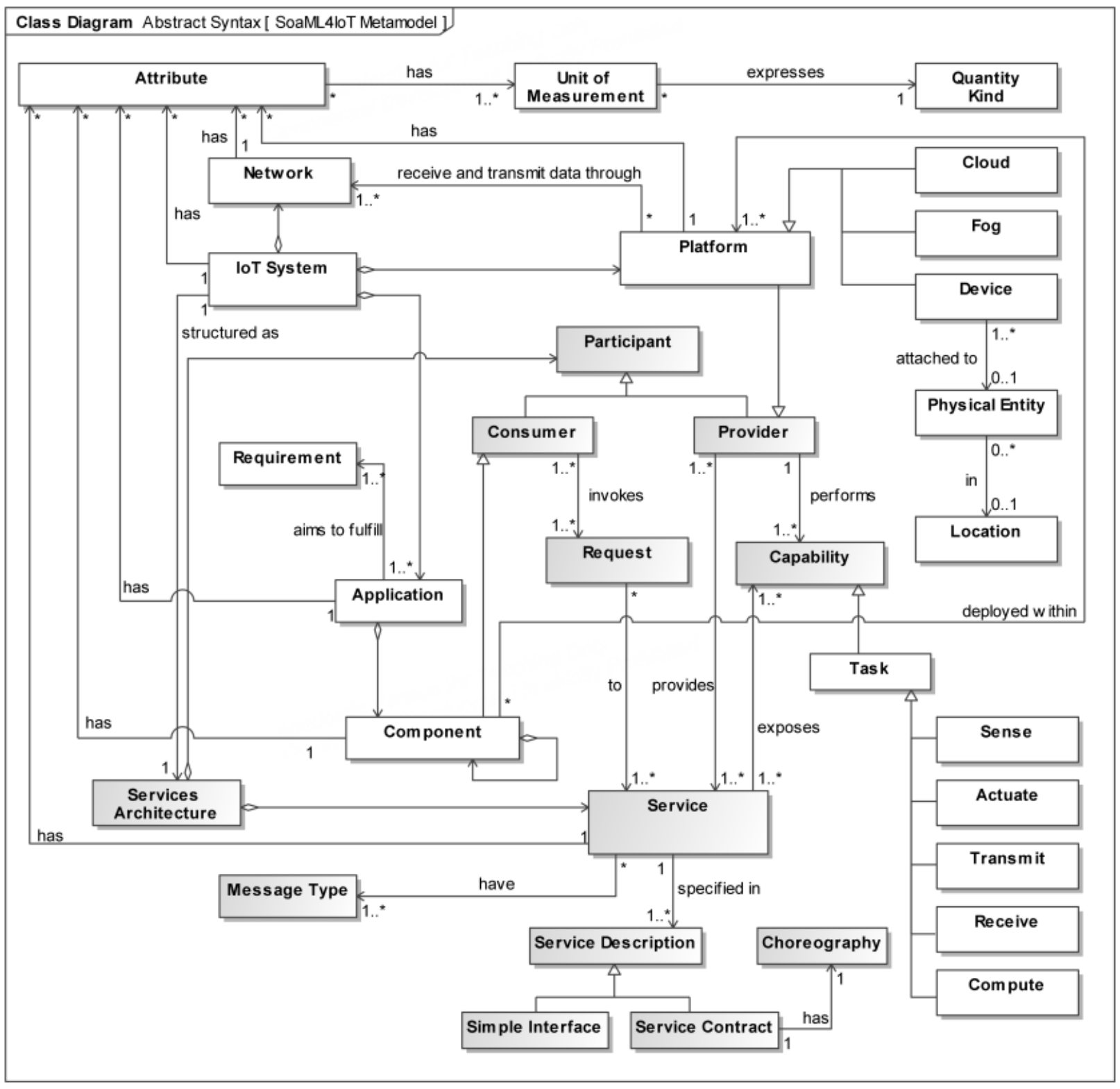

Figure 2. The SoaML4IoT Metamodel

The devices enhance physical entities with sensing, actuating, communication, and computing capabilities. *Sense* tasks aim to collect data from physical entities (e.g., car speed, human body's temperature) or the environment it is inserted into (e.g., room's temperature or lighting level); *actuate* tasks can affect the physical realm (e.g., turn on/off a heater, sound alarm); *transmit* and *receive* tasks regard the communication of the device; finally, the *compute* task refers to the processing capability of the device. The platforms communicate with each other and eventually with the Internet by using wired or, more often, wireless networks. A *Network* refers to the set of nodes and links providing the communication path by with the platforms receive and transmit data. An *Application* refers to a set of specialized algorithms that request services and process data aiming at fulfilling user-defined requirements. An application is composed of one or more *Components*, which are software units that can be deployed within platforms. Note that the metamodel represents the IoT applications as a set of components instead of a monolithic structure.

In SoaML4IoT metamodel, the capabilities performed by the providers are specialized as tasks (sense, actuate, transmit, receive, and compute). In turn, the providers are specialized as platforms. Components act as consumers, requesting the required tasks exposed through the devices' services. In other words, devices attached to physical entities expose their provided tasks through services, which are consumed by application's components aiming at fulfilling the stakeholders' requirements.

### 4.2 Concrete Syntax

The concrete syntax of SoaML4IoT (Figure 3) consists of an UML profile extending the UML elements Port, Class, Dependency, Interface, Association, and DataType. Such extensions realize the concepts with the same name from the SoaML4IoT metamodel. Instead of specializing the stereotypes of SoaML, we decided to extend the elements from the UML with concepts of our metamodel. Note that we extend the same UML elements as the SoaML. For example, the concept *Consumer* from the SoaML metamodel extends the UML element Class. Thus, the concept *Application* from SoaML4IoT metamodel, which specializes consumer, extends the same UML element Class. The reasoning behind this design decision is that fUML engines do not handle cases were multiple stereotypes are applied over the same modeling construct [43]. Thus, if we specialize the SoaML profile, the same class could be applied with multiple stereotypes (e.g., provider and device), hampering the model execution. Another adaptation we made to allow the execution regards the *Service Contract* and *IoT System* elements (this later specializing the concept of *Services Architecture*). In SoaML, both service contract and service architecture extend the UML element *Collaboration*, which is not part of the fUML semantic model. Thus, we extended the UML element Class, which also allows the modeling of collaboration through composite structures.

Following the second principle of DSL's, each stereotype of SoaML4IoT provides properties to specify the general properties of the elements composing the solution. In other words, the elements do not provide properties that are onçy relevant for a specific expert. For example, a device specialist may be interested on the *processing speed* of the devices. On the other hand, a network specialist may be interested on the *network interface*s available for the platforms; or the software architect may be concerned on the *execution environments* (Python, Java) provided by the platforms. Although processing speed is an essential characteristic for the device specialist, it may not be relevant for the network specialist. The same may happen with network interface and execution environments. Therefore, instead of representing every aspect of an IoT system, of interest for every possible stakeholder, as presented in Section 4.4, the framework is fully extensible, so that the experts can add new attributes according to their specific needs [1].

The *Service* has the attribute *protocol*, aiming to specify its application-level protocol (e.g., HTTP, CoAP). The *Application* also has the attributes *region* and *localTimer*. The former aims to specify the geographic location of the target environment (as requested by the application), from which the sensing data will be collected and/or actuation tasks will be performed upon. The second attribute is used by the execution semantics as a counter, considering the periodic applications. Thus, for example, when the user specifies that the application requests a given service every 2 minutes, the attribute *localTimer* is used to control the time elapsed since the last request (such behavior is specified in the execution semantics, introduced in the section).

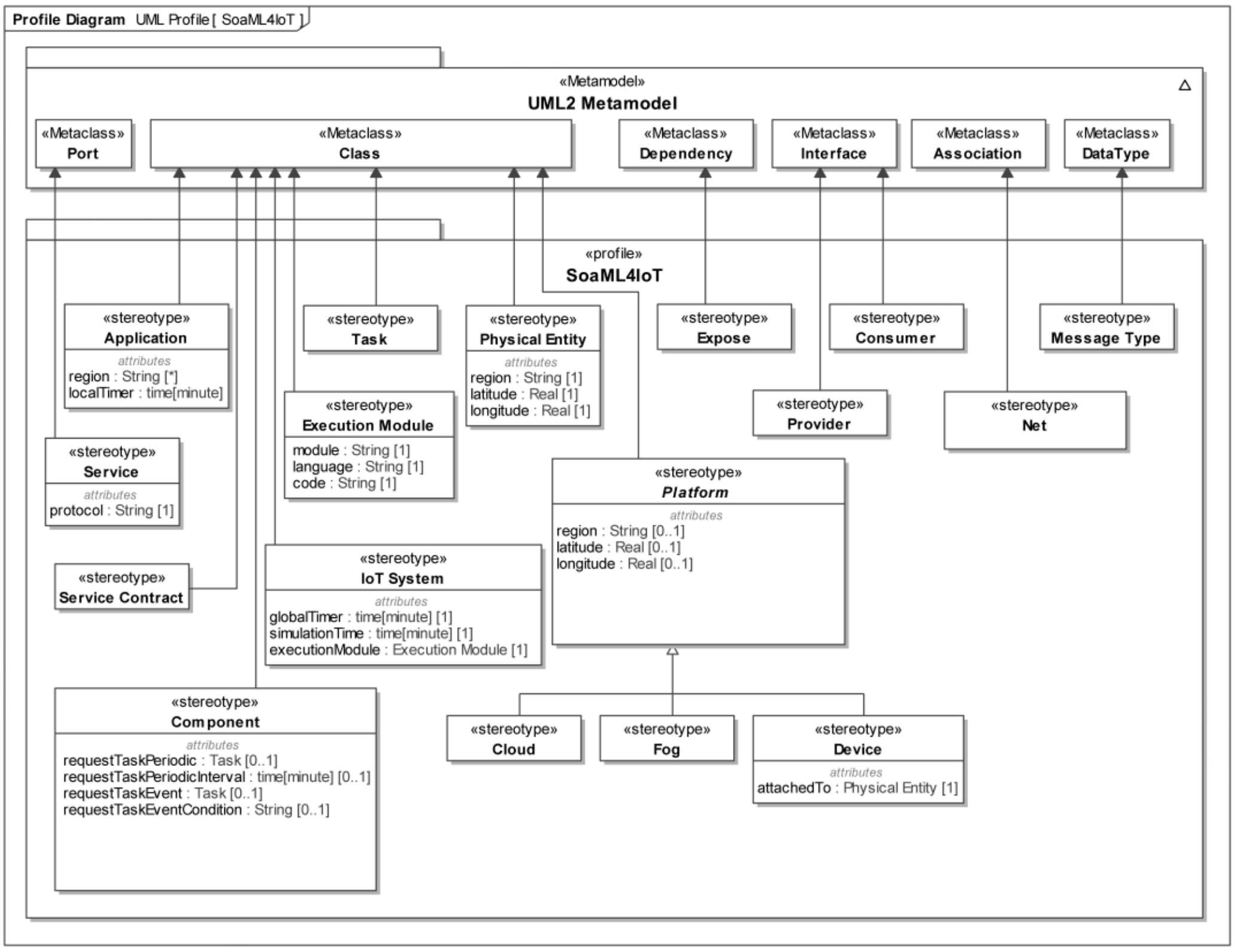

Figure 3. The SoaML4IoT UML Profile

The *Component* stereotype has the attributes for specifying periodic and/or event-based applications. Recall that an application is composed of one or more components. Thus, although the periodic or event-based model refers to the application, the components are the responsible for actually performing that required requests. The attribute *requestTaskPeriodic* models the task to be requested at regular time intervals, which is specified in the attribute *requestTaskPeriodicInterval*. In the simulation, such an attribute is used in conjunction with *localtimer* to trigger the request. On the other hand, the attribute *requestTaskEvent* defines the required task that aims to be requested when an event occurs. Such an event is specified in SoaML4IoT textually, in the attribute *requestTaskEventCondition*.

---

[1] It is important to highlight that SoaML also adopts this strategy, as introduced in its specification [26]: *SoaML focuses on the basic service modeling concepts, and the intention is to use this as a foundation for further extensions both related to integration with other OMG metamodels like BPDM and BPMN 2.0, as well as SBVR, OSM, ODM, and others.*

The *Physical Entity* stereotype allows specifying its geographic *region* through the *latitude*, *longitude* attributes. The *Platform* (abstract) also allows specifying its geographic *region* through the *latitude*, *longitude* attributes. The Device stereotype has only *attachedTo* attribute, which models the physical entity equipped with the device.

Finally, the *IoT System* stereotype has three attributes, *globalTimer*, *simulationTime*, and *executionModule*. All these attributes are used in the model execution. The attribute *globalTimer* represents the simulation time in minutes, which is updated until it achieves the required simulation time (*simulationTime* attribute). At the beginning of the simulation, it is initialized with 0 (zero), and it is incremented until achieving the limit defined by the user or the user stops the simulation. The *Execution Module*, which is the third attribute of the IoT system, is the extension point of the execution semantics. It has three attributes, namely, *module*, *language*, and *code*. The first specifies the module name. The second defines the language used to implement the module. Finally, the third attribute is used to indicate the location of the code file to be executed in the simulation. The details about the extension of the execution semantics of SoaML4IoT through the execution module is explained in the following.

### 4.3 Execution Semantics

The details and implementation of the execution semantics of IoTDraw are available at brccosta.github.io/iotdraw. The current version of the execution semantics is applied over the *IoT System* stereotype, which models the behavior of a periodic IoT application. Thus, it represents a discrete-event system [59], in which the service requests occur at a discrete set of point in time, as defined by the data sending interval (period of time between two consecutive data sending) required by the application (which is modeled in the *requestTaskPeriodicInterval* attribute).

List 1. Execution Semantics of IoT System (Language: Alf)

```
1  activity execute(){
2   WriteLine(“[IoT System Execution]”);
3   WriteLine(“Begin...”);
4   UML::Stereotype execModule = this.types.getAppliedStereotype(“Execution Module”);
5   if(executionModule != null) {
6     String module = execModule.getAttributeValue(“module”);
7     String code = execModule.getAttributeValue(“code”);
8     String language = execModule.getAttributeValue(“language”);
9     WriteLine(“Execution module: ” + module);
10    /*@inline(language)
11    String result = module(this);
12    */
13    WriteLine(result);
14  }
15  UML::Stereotype iotSystem = this[0].types.getAppliedStereotype(“IoT System”);
16  Integer globalTimer = iotSystem. getAttributeValue(“globalTimer”);
17  Integer simulationTime = iotSystem.getAttributeValue(“simulationTime”);
18  List<Application> applications =
19  this.types.getAppliedStereotype(“Application”);
20
21  //Perform the simulation
22  while(globalTimer <= simulatiomTime){
23   for(IoTDraw::Application app in applications){
24    for(IoTDraw::Component comp in app.types.getAppliedStereotype(“Component”);
25     if(comp.getAttributeValue(“requestPeriodicInterval”)== app.getAttributeValue(“localTimer”) &&
        comp.getAttributeValue(“requestTaskPeriodic”)!= null)){
26      UML::Stereotype contract = this[0].types.getAppliedStereotype(“Service Contract”);
27      contract.execute();
28      app.setAttributeValue(“localTimer”, 0);
29     } else {
30      Integer localTimer = app.getAttributeValue(“localTimer”);
31      localTimer++;
32      app.setAttributeValue(“localTimer”, localTimer);
33     }
34   }
35   globalTimer++;
36  }
37 }
```

In List 1 , the first step of execution extension is to verify whether there is any execution module assigned to the IoT system class (lines 4 and 5). If it is true, the module code is injected as an opaque action, which is responsible for executing such a module (line 11). Such an injection is performed by IoTDraw at runtime. In line 16, we get the *globalTimer* tagged value, which is 0.0 by default, and in line 17 we get the required simulation time. This simulation time is configured by the modeler and represents the total time required by the simulation in minutes. In line 18, we get all the applications that comprise the IoT system. It is important to highlight that, in this way, we allow the simulation of various applications comprising an IoT system.

The simulation itself is performed into a while loop, which starts at line 22. Each iteration of the loop represents a minute in the simulation. The condition of the loop is that the simulation time (specified in the attribute *simulationTime*) is less or equals to the simulation time (specified in the attribute *simulationTime*). Thus, the simulation runs until it achieves the time limit specified by the modeler. Inside the while loop, the logic is to verify all components of all applications of the IoT system (lines 23 and 24). Whether the component requires a task periodically (line 25), such a component is a candidate to be executed. Another verification that we do in line 25 is whether the local timer equals the *requestTaskPeriodicInterval*. As introduced in Section 2.2, the local timer attribute is an internal counter of the application, and the *requestTaskPeriodicInterval* defines the data sending rate of the component. By comparing it with the local timer, we ensure that the component only will be executed when it is required. Next, in line 27 we execute the choreography that is related to the current object. After executing the choreography, the local timer attribute is re-assigned to 0.0. Line 29 is executed whether the local timer does not equal to the periodic interval required by the application; thus, we only increment the local timer. In line 34, we also increment the global timer.

### 4.4 Extensibility Rules

Following the third and fifth principles of language engineering, SoaML4IoT can be extended by following the extensibility rules. The extensibility rules of SoaML4IoT are based on the techniques of Model-Driven Architecture (MDA) [60], that is, (i) creating new attributes (i.e., tagged values) for the existing elements of the language (i.e., the stereotypes of the SoaML4IoT profile), (ii) specializing existing elements, and (iii) implementing user-defined execution modules to be called in the model execution. Extension for the SoaML4IoT that follows these rules keep the language compliant with the metamodel and the execution semantics. In this way, modelers can enhance the expressivity of the language by providing specific attributes based on the domain of interest. On the other hand, user-defined execution modules provide means to simulate other application scenarios and, thus, support answering several design questions. Creating new attributes for existing elements of SoaML4IoT follows the same methodology of UML profiling.

The execution modules can be implemented using standard object-oriented mechanisms, that is, inheritance and polymorphism. To create the execution module, the user must implement the interface *ExecutionModule*, as detailed in List 2.

List 2. ExecutionModule Interface

```
interface ExecutionModule{
 String executionModule(IoTSystem iotSystem);
}
```

The function *executionModule* receives as argument an object typed of *IoTSystem* and responds a String data. When a class implementing such a module is called by the IoTDraw framework, it collects the UML Class stereotyped as IoT System with all its composing elements, that is, Platforms (Cloud, Fog, Device), Applications (and composing Components), and Networks. IoTDraw structures such an object as a POJO (Plain Old Java Object) class [61]. POJO classes refer to pure data structures that has fields with getters and setters. The tagged values are converted into class attributes and their types converted into basic types. For example, *time[minute]* (type of *globalTimer* tagged value) is converted into Double.

### 4.5 Model Execution Engine

Aiming to apply the fourth principle of language engineering, the execution semantics specified above has been developed as a prototype on Moka [30], an Eclipse Papyrus module that includes an execution engine complying with fUML. When executing a SoaML4IoT model, the extended engine analyzes each element. According to the stereotype applied over the element, the code representing the execution semantics of the element is dynamically injected at runtime to reflect the execution semantics applied by the stereotype. Figure 4 depicts the components of our proposal focusing on the extension mechanisms of SoaML4IoT.

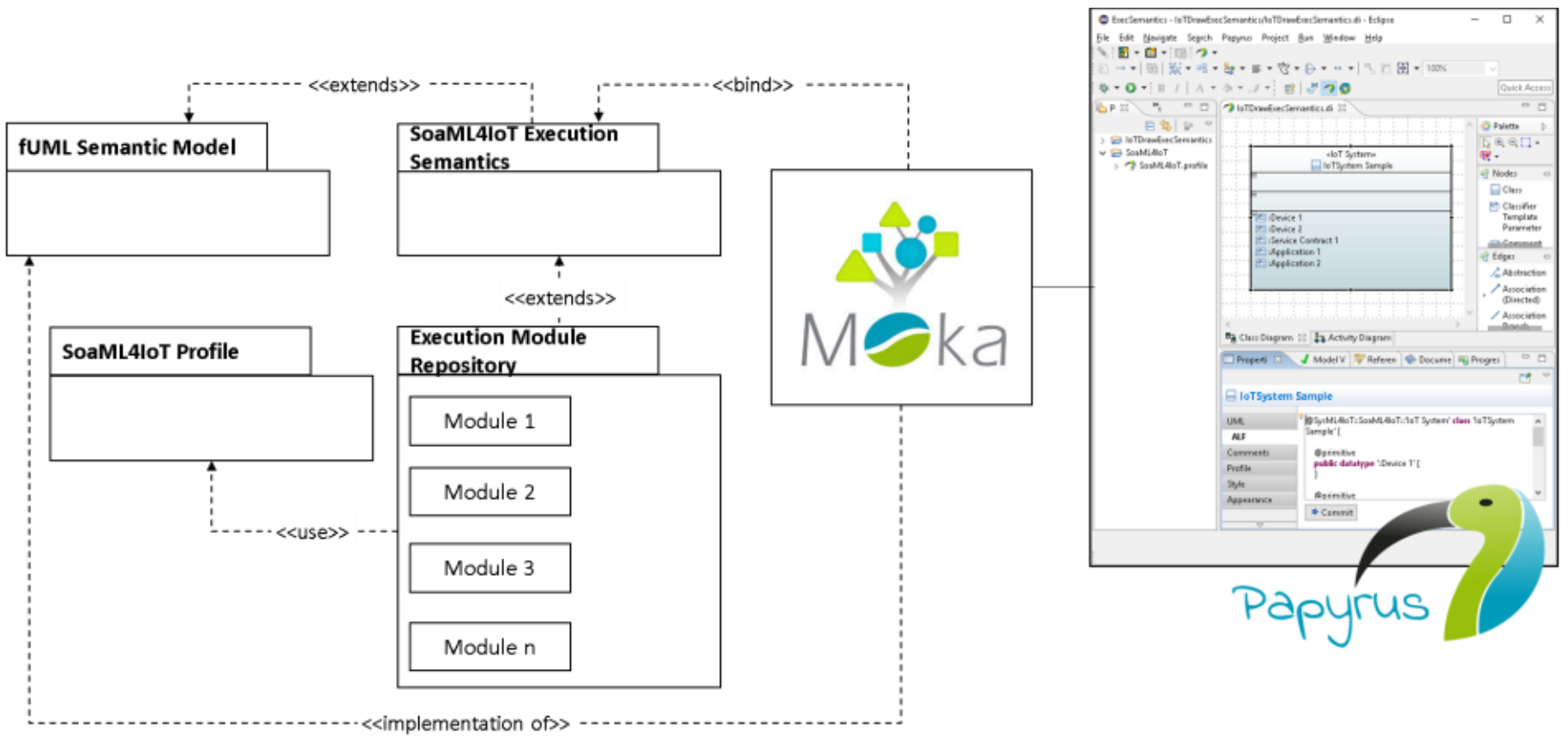

Figure 4. IoTDraw framework – Model execution semantics and extension modules

## 5 Evaluation

In this Section, we evaluate our proposal based on two methodologies, namely proof-of-concept (PoC) and the perceived acceptance of IoTDraw by software engineers from industry. In the PoC, we show the specification of the FW application introduced in Section 3, highlighting the support of our framework to answer important design questions. The second evaluation study is based on the Technology Acceptance Model (TAM) [62], a mature theoretical model that has been widely used in many empirical researches.

### 5.1 Evaluation Proof-of-Concept: FW Application specification

Returning to the running example introduced in Section 3, on the PSC infrastructure (Figure 1), we want to design IoT applications aiming to enhance the quality of the public services offered to citizens. Due the lack of space, we focus on the specification of the Flood Warning (FW) application (the complete specification of all applications is available at brccosta.github.io/iotdraw). The objective is to sense the water level at regular time intervals and send alerts to the citizens, in case of flooding, to enable the fast evacuation near the flooding area. The FW application also performs analytics aiming at examining different flooding scenarios, which will be helpful for urban planning. The FW application models, which will be created by SoaML4IoT, must support modelers to (i) identify key elements of the system, and (ii) answer important design questions regarding the services approach as well as issues related to lifetime, availability, and response time QoS requirements.

As identified by the literature, IoT applications have highly heterogeneous requirements, and can vary from being composed of only a few devices, such as e-health application ([63]), to hundreds and even thousands of heterogeneous devices and software components (e.g., [64]). In our example, we specify the FW application with a limited number of elements in such a way that it would be possible to explain our framework in detail. Anyway, applying our framework to specify and analyze more complex IoT applications follow the same approach that we use to model the running example. In Section 5.2, after introducing IoTDraw and exemplifying it with the FW application, we modify its architecture, resulting in a more complex scenario that reinforces the need and usefulness of our proposed framework.

The specification of the FW application follows the process for service-oriented design proposed by Erl [65]. The diagrams created in each step follow the SoaML approach for structuring the modeling components. In summary, the architecture specification of an SOA-based IoT application using SoaML4IoT consists of three kinds of models: (i) Service Contract and Choreography (Figure 5 and Figure 6); (ii) Participants (Figure 7 and Figure 8); and, (iii) Services Architecture (Figure 9). We provide a briefly introduction of each model in the following. A more detailed explanation about the process and the SoaML approach can be found at [65] and [26].

The Service Contract models specify the constraints and design standards the provider and consumer participants must adhere when interoperating such as the protocol and the structure of the information that is transmitted from the provider to the consumer. The contract also defines the behavior of the service's communication; that is, the actions that are performed by each participant as well as the order in which these actions must be performed. Technical issues related to software components and hardware devices are specified in the Participant's models. As introduced in Sections 3 and 4, SoaML4IoT adopts the 3-tier architecture model, which encompasses computing nodes that can be of type Cloud, Fog or, Device. In the Participant Provider models, we specify the tier each node is part of. Finally, Services Architecture models specify the fundamental structure of the system by composing the software and hardware elements. The composition of each element is mediated by the service contracts specified in the service contract models. By mediating the composition through contracts, both provider and consumer participants must agree with the constraints and design standards when composing the architecture, allowing the interoperability between all elements of the system.

The approach for the specification of FW application adopts the concept of component-specification and component-use [21]. This approach stablishes that in the development lifecycle the components are conceived to be reused, forming component library or repository. Thus, when designing a system's architecture, the modelers take components from a repository and assembles them into a system. In the case of the FW application, for example, the consumer and provider participants are specified in the Participant Provider models and are used in the Services Architecture specification; the same happens with the contracts, which are specified in the Service Contract models and used in the Service Architecture specification.

#### 5.1.1 Service Contract Design

There are two contracts that the participants must agree when designing the consumers and providers of the FW application, namely, *Request Water Sensor* and *Request Alarm* (Figure 5). The *Monitor Water* task provides two operations: *sense*() and *transmit*(). The former aims at sensing the water level while the second transmits the data to the requesting component. The data is structured as the *WaterLevelData* message type. The *Monitor Water* task is exposed by the *FW_SenseInterface*, which is realized by the *Water Sensor* (provider interface). This interface is used by *Flood Monitor* (consumer interface), which, in turn, realizes the conjugated interface *~FW_SenseInterface*.

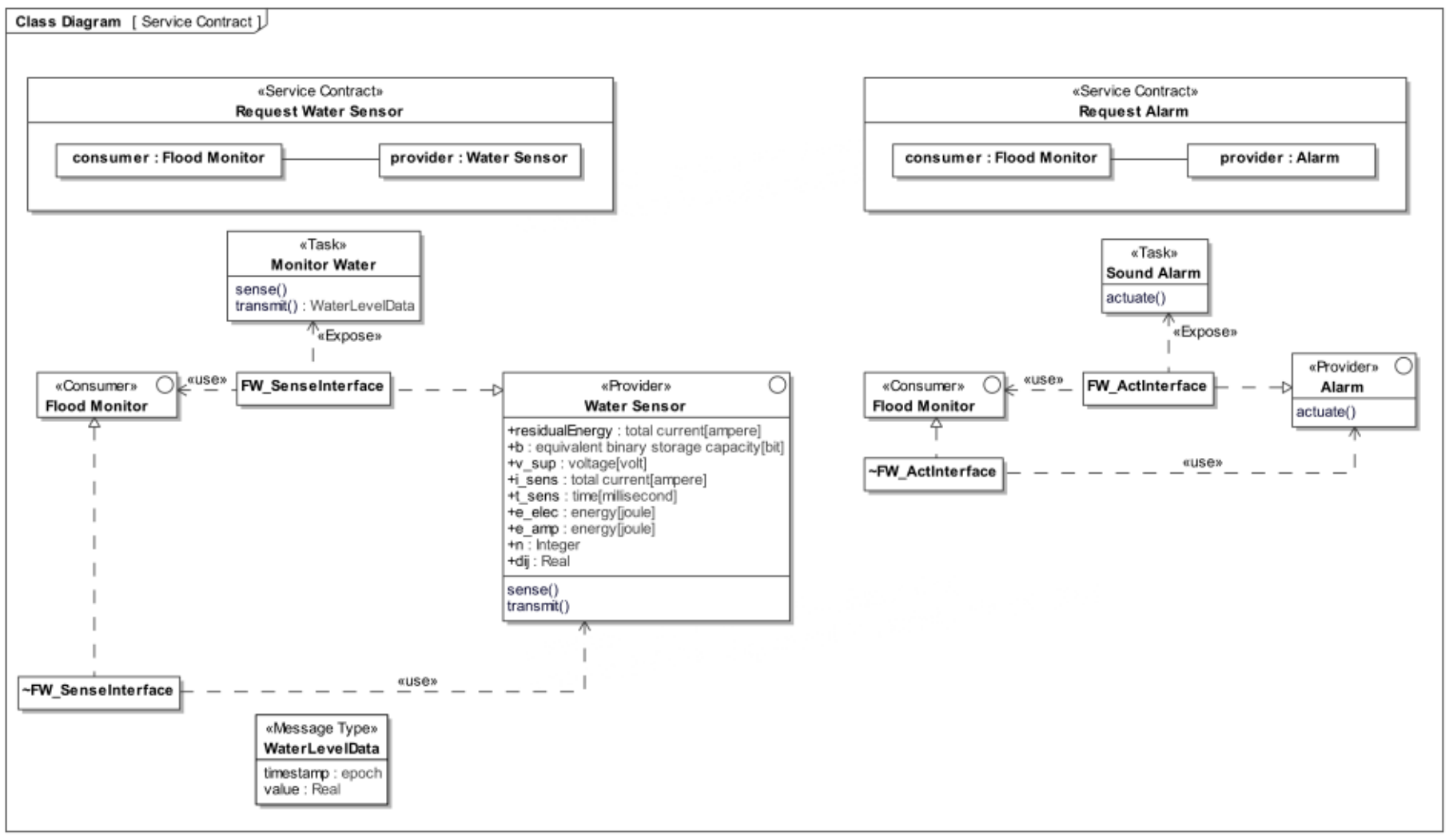

Figure 5. Contract Design for FW Application

In the *Request Alarm* service contract, the *Sound Alarm* task provides only the operation *actuate*(), which aims to sound the alarm in case of flooding. The Sound Alarm task is exposed by the *FW_ActInterface*, which is realized by the *Alarm* (*Provider* interface). This interface is used by *Flood Monitor* (consumer interface), which, in turn, realizes the conjugated interface ~*FW_ActInterface*.

In the *Request Water Sensor* service contract, the *Water Sensor* (provider interface) specifies a set of properties that a participant implementing such interface must provide. The devices are powered by limited batteries; thus, these properties are required to answer the following design question (**DQ1**) "*considering the data sending rate required by the application, what is the operational lifetime of the Water Sensor devices?*". Recall that, in the context of the PSC, the lifetime is considered as the time spanning from the instant when the device starts functioning until it runs out of energy, becoming the service provided by such a device unavailable.

To answer this design question, we adopt an analytic model proposed by Halgamuge and colleagues [66] to predict the energy consumption of devices' tasks. In such a model, the energy consumption of sensing tasks $E_{sense}$ (unit: Joules) is given by:

$$E_{sense} = bV_{sup}I_{sense}T_{sense}$$

where $b$ is the bit package (unit: kilobit - kb) collected by the sensing activity, $V_{sup}$ is the supply voltage (unit: volt - v), $I_{sense}$ is the total current (unit: milli ampere) required for sensing activity, and $T_{sense}$ is the time duration (unit: milliseconds - mS) for sensing unit is collecting data from the environment. The equation from Halgamuge's model formalize the energy consumption of transition task $E_{transmit}$ (unit: Joules), is given by

$$E_{transmit} = bE_{elec} + bd_{ij}^{n}E_{amp}$$

where $b$ is the bit package to be transmitted in a distance $d_{ij}$ (unit: meter - m), $E_{elec}$ (unit: Nano Joules per bit – nJ/bit) is the energy dissipated to transmit data, which may differs considering the network interface, for example; $E_{amp}$ is the energy dissipated by the power amplifier (unit: Pico Joules per bit per square meter – pJ/bit/m$^2$), and $n$ is the distance-based loss exponent (unit: Integer).

The battery capacity of devices is typically measured as milliampere hour (mAh), while the energy consumption is measured as Joules (J). Thus, it is necessary to convert from J to mAh in order to verify the difference between the battery capacity and the total energy consumption required for the device. The conversion from J to mAh is given by:

$$Joules_to_mAh = \frac{1000 \times E_{(Wh)}}{V_{(V)}}$$

where $E_{(Wh)}$is the energy in watt-hours (Wh) and $V_{(V)}$is the voltage in volts (V). 1 J corresponds to 0.000277778 Wh, thus $1\,J = 1000 \times 0{,}000277778/V_{(V)}$ mAh.

These energy consumption equations are modeled in the choreography of the *Request Water Sensor* service contract, specified as an fUML Activity Diagram (Figure 6). It starts with a *readSelf* action followed by the reading of the attributes (*readStructuralFeature* action) required to calculate the energy consumption of sensing task. An *opaque action* calculates the energy consumption. The result in Joules is converted with an opaque action into mAh (milli ampere hour), which is the unit of battery capacity of devices. Next, this result is used to update the residual energy. The behavior of the transmit task follow the same structure.

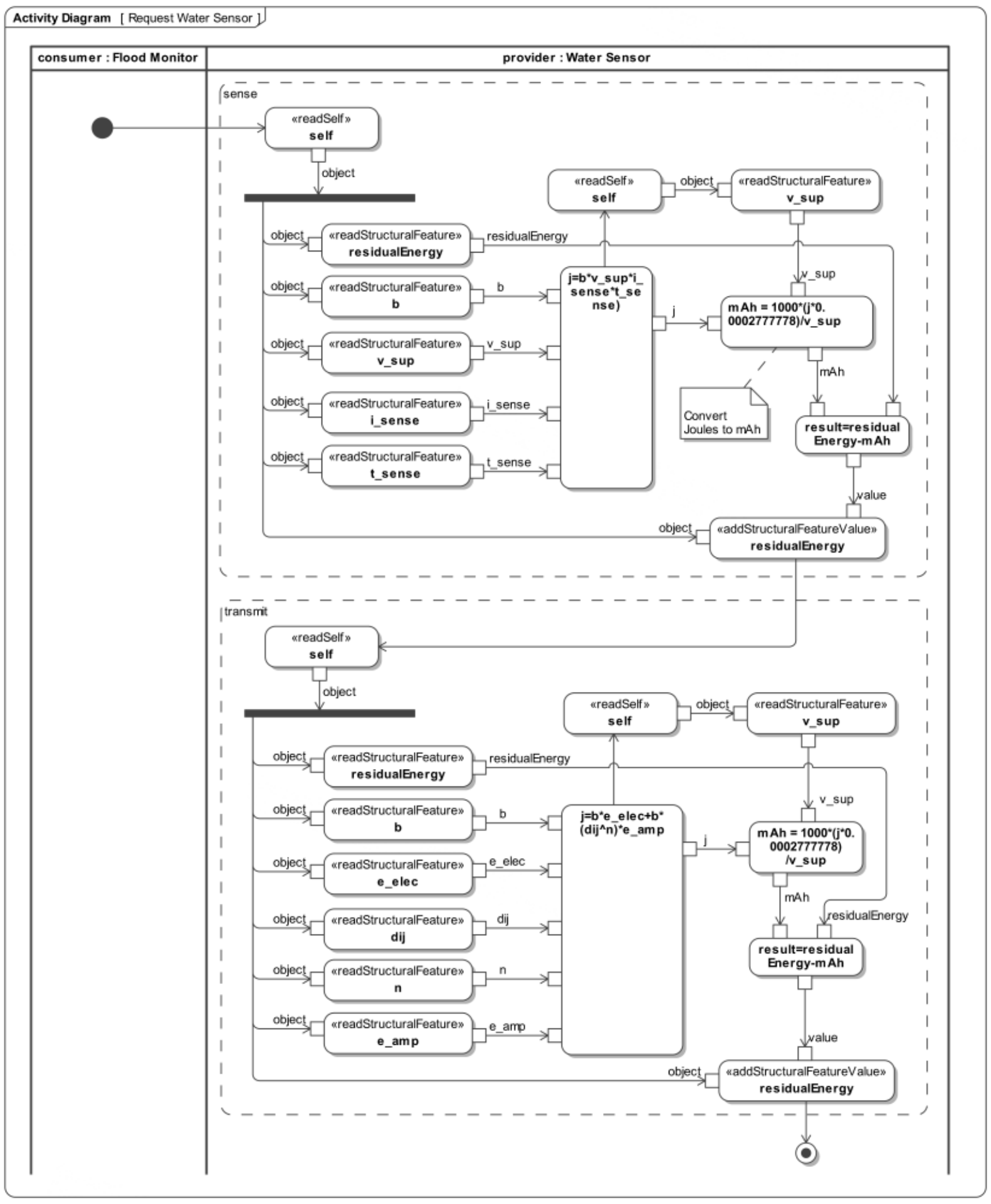

Figure 6. Choreography of Request Water Level Sensor Service Contract (Key: fUML Activity Diagram)

### 5.1.2 Participants Design

The specification of consumer participants of FW application is depicted in the UML Class Diagram of Figure 7. The application Flood Warning aims to monitor the water level in two important avenues of Padova, namely, Via Codalunga and Via Niccoló Tommaseo. The application is composed of three components, namely, *FloodAPI*, *FloodMonitor*, and *Analytics*. The component *FloodAPI* intermediates the communication between other components and the devices' services. It has two services; each one typed as an interface as defined in the respective contract; thus, the *FloodAPI* requires the *Alarm* and the *Water Sensor* interfaces. The component also provides a service, based on HTTP, providing an interface for external requests.

The component *FloodMonitor* is responsible for managing the monitoring for flooding. It requires the interface *Flood API* and provides a service, based on HTTP, with an interface *Flood Monitor* for external requests. The task that will be requested periodically by the component (i.e., every 2 minutes), is the *Monitor Water*. In turn, the task *Sound Alarm* will be requested only when the water level is above 20 centimeters. Finally, the component *Analytics*, which requires the interface *Flood Monitor*, aims to store the water level data and further perform analytics aiming at examining different urban scenarios. This component requires the interface Flood Monitor.

The specification of provider participants of FW application is depicted in the UML Class Diagram of Figure 8. There are two cloud nodes available for the system; the first is in *Stuttgart* while the second is in *Michigan*. The three available fog nodes, *fog_1*, *fog_2*, and *fog_3*, are located in different areas of the city of Padova. Finally, in the example, there are two devices, namely, w*ater sensor* and *alarm*. The devices have services providing interfaces for external access. Note that these services are typed as elements of the service contracts.

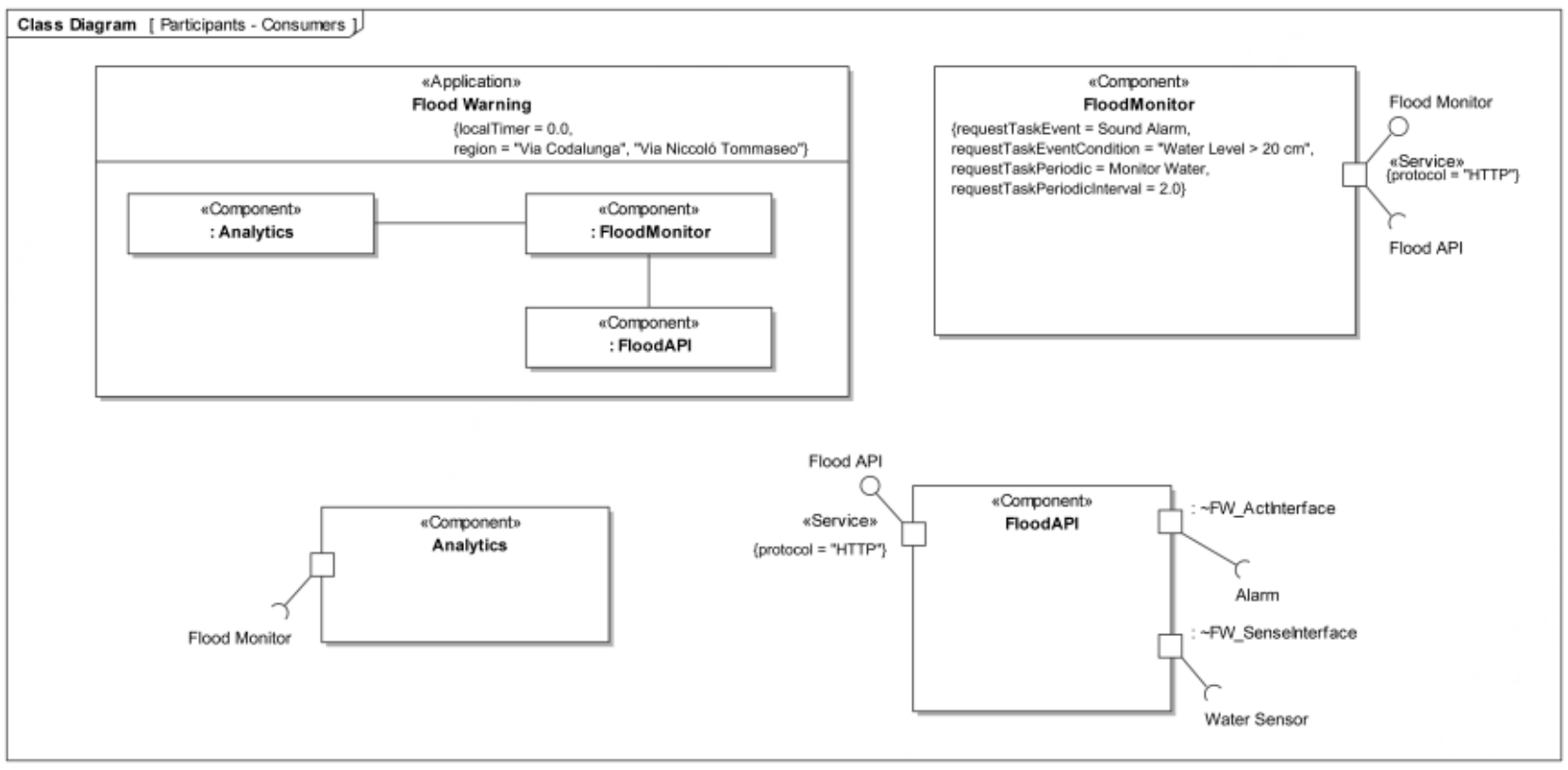

Figure 7. Participants of FW application – Consumers

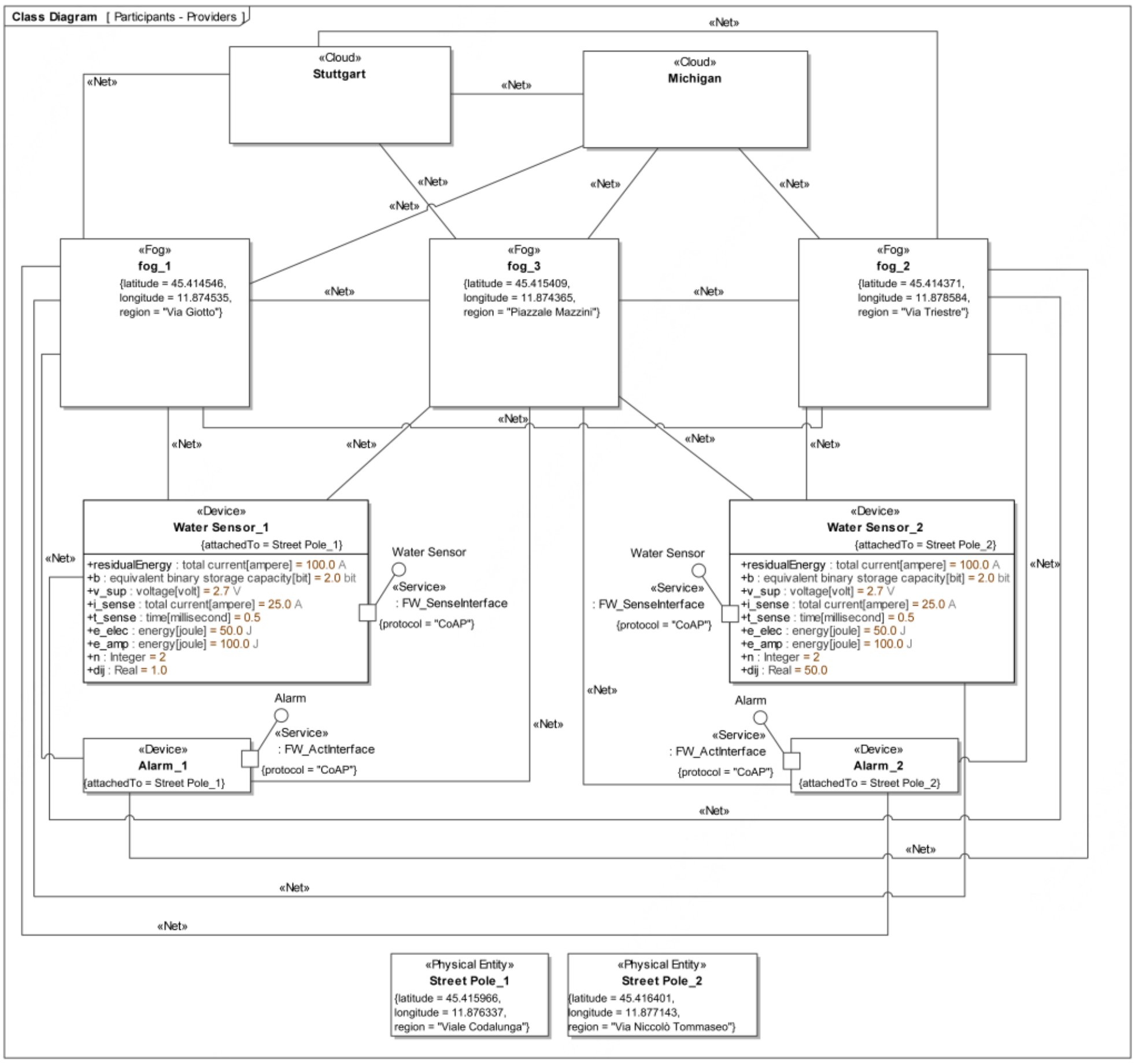

Figure 8. Participants of FW application – Providers

Since the water sensor devices realize the *Water Sensor* interface, it provides the properties required by this interface (recall that these properties are used in the choreography to estimate the energy consumption of the services). In our example, we use the values of voltage, current, etc. presented by [66], considering a generic device (e.g., Arduino, RaspberryPi). Finally, the associations stereotyped as <<Net>>, models the network connections between the platforms.

### 5.1.3 Services Architecture Design

The services architecture design of PSC, including the FW application, is depicted in the UML Class Diagram of Figure 9. The PSC services architecture aims at connecting the consumers (i.e., applications) to the providers (i.e., devices) through the specified contracts. When connecting the participants, IoTDraw checks if the applications have components that agree with the contracts, that is, if they have the ports implemented the required interfaces as specified in the contracts. It is also specified the simulation time of the model, that is, 1.051.200 minutes (i.e., 2 years).

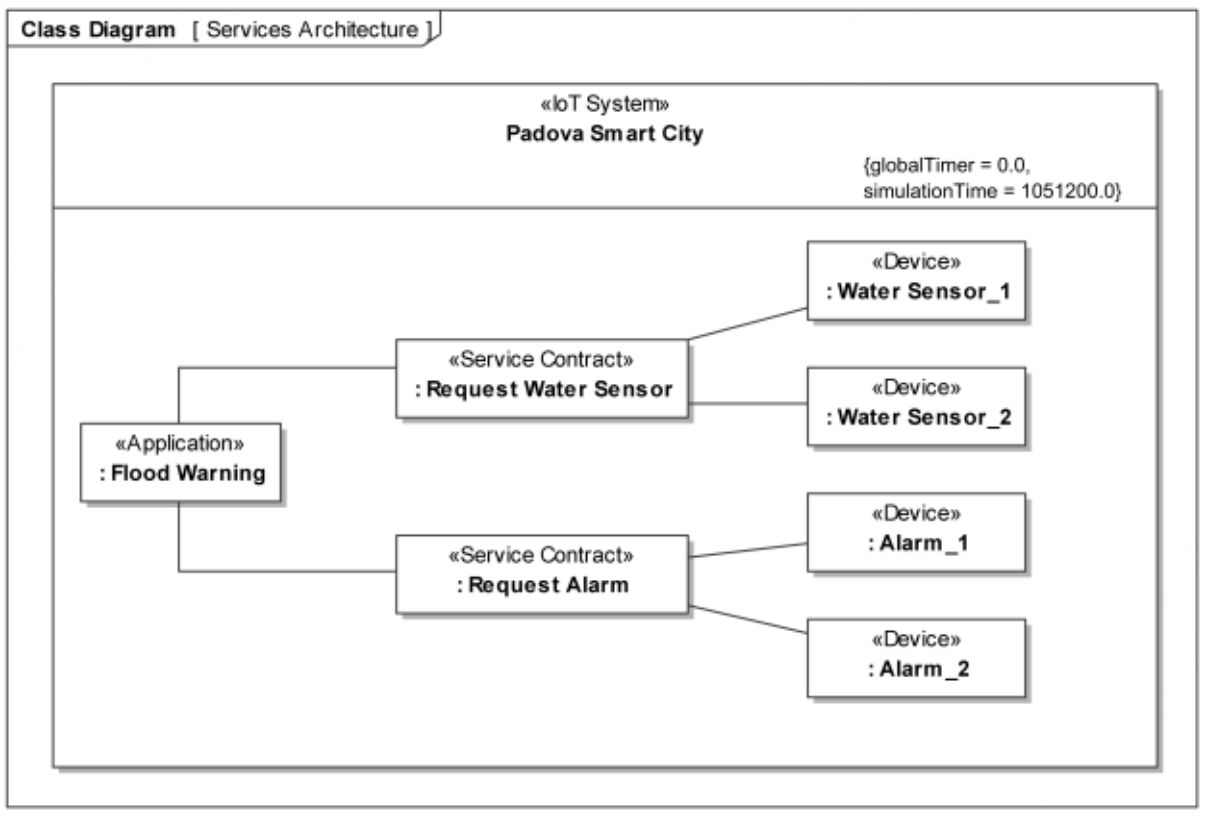

Figure 9. The services architecture of PSC System

### 5.1.4 Answering Design Questions

After specifying the FW application, there are important design questions (**DQ**) that must be answered regarding lifetime, data freshness, availability, and response time QoS requirements. Some of these questions are: (**DQ1**) *considering the data sending rate required by the application, what is the operational **lifetime** of the Water Sensor devices?*; (**DQ2**) *in the $S^2$aaS model, what is the operational lifetime of the devices considering the **data freshness** required by service consumers?*; (**DQ3**) *in which type of node (cloud and/or fog) should the application's components be deployed?*; (**DQ4**) *what is the deployment configuration that promotes the highest **availability** for the FW application?*; (**DQ5**) *what is the deployment configuration that promotes the lowest **response time** for the FW application?* By answering such questions, the modelers would make more successful architectural decisions about, for example, deployment configurations, component replication, protocol specifications, and caching.

Based on the extensibility rules introduced in Section 4.4, we developed the execution modules depicted in Figure 10 to support the modelers answering the design questions presented above. Furthermore, we extended the language to represent the required properties. The models created with the extended version of SoaML4IoT are depicted in Figure 11 and Figure 12. All modules with their complete specifications are also available at brccosta.github.io/iotdraw.

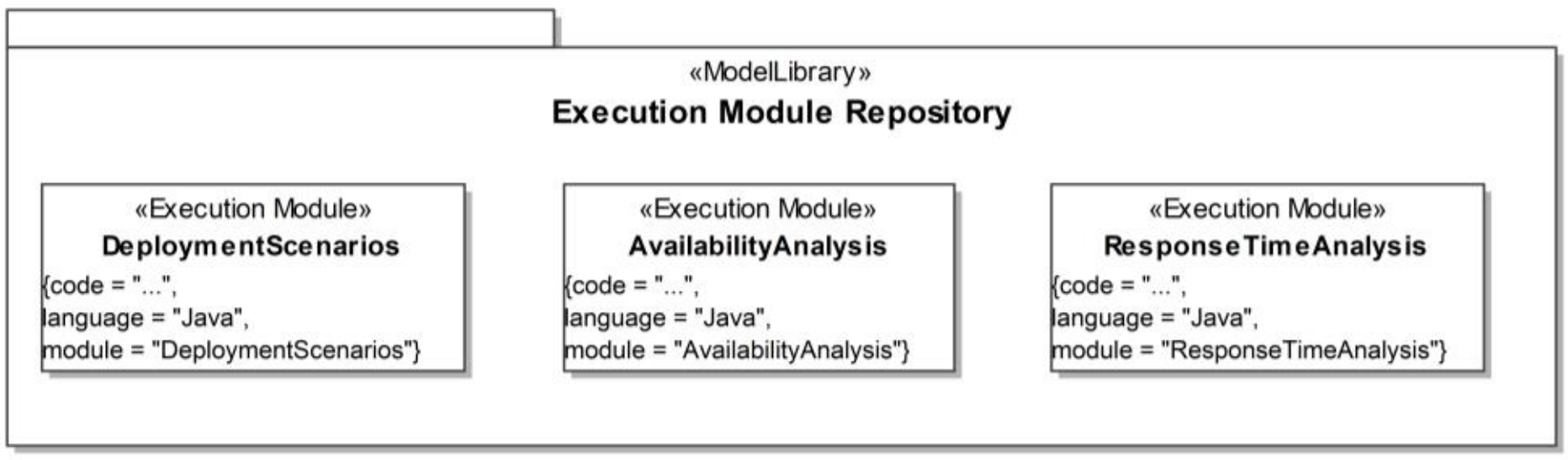

Figure 10. The execution module repository

The *DeploymentScenarios* module supports finding the best deployment configuration for the application modules. When deploying a component within a platform, it is necessary to verify whether such a platform provides the software capabilities that are required by the component. For example, whether a component was developed by using C#, to allow executing it, the platform must provide the .NET framework, which is the execution environment of C# applications. To allow the analysis, the modelers must inform the required and provided software in the components and platforms, respectively.

The *AvailabilityAnalysis* module extends the *DeploymentScenarios* module by allowing to find the deployment configurations with the highest availability. To achieve this goal, the module apply the following availability model [15]:

$$Availability(P) = \frac{MTBF}{(MTBF + MTTR)}$$

where $P$ is the platform, $MTBF$ refers to the mean time between failures and $MTTR$ refers to the mean time to repair. The modelers must provide such information in the platforms.

Finally, the *ResponseTimeAnalysis* also extends the *DeploymentScenarios* module by allowing to find the deployment configurations with the lowest response time between components. To achieve this goal, we apply the following response time analytic model [56]:

$$ReponseTime(R) = L(Network) + P_{time}(Platform, Component)$$

where $R$ denotes the request. $L$ denotes the average latency of the network connection between consumer and provider participants. And, $P_{time}$ refers to the mean processing time of the platform to process the component's algorithms. And, $Q$ denotes the waiting time in the queue. $P_{time}$ is given by [21]:

$$P_{time}(Platform, Component) = \frac{CPU_{demand}(Component)}{CPU_{frequency}(Platform)}$$

where the average CPU demand of the component (unit: cycles) is divided by the CPU frequency (unit: GHz) of the Platform. All such information must be provided by the modelers in the platforms and components.

We execute the model aiming to answer the design questions (DQ1 to DQ6) introduced earlier. In the analysis, we execute the model in Intel Core i7-2640M, 2.80GHz, 6GB of RAM, JRE version 1.8.0_191. The execution follows the semantics introduced in Section 4.3. In the following, we present the results.

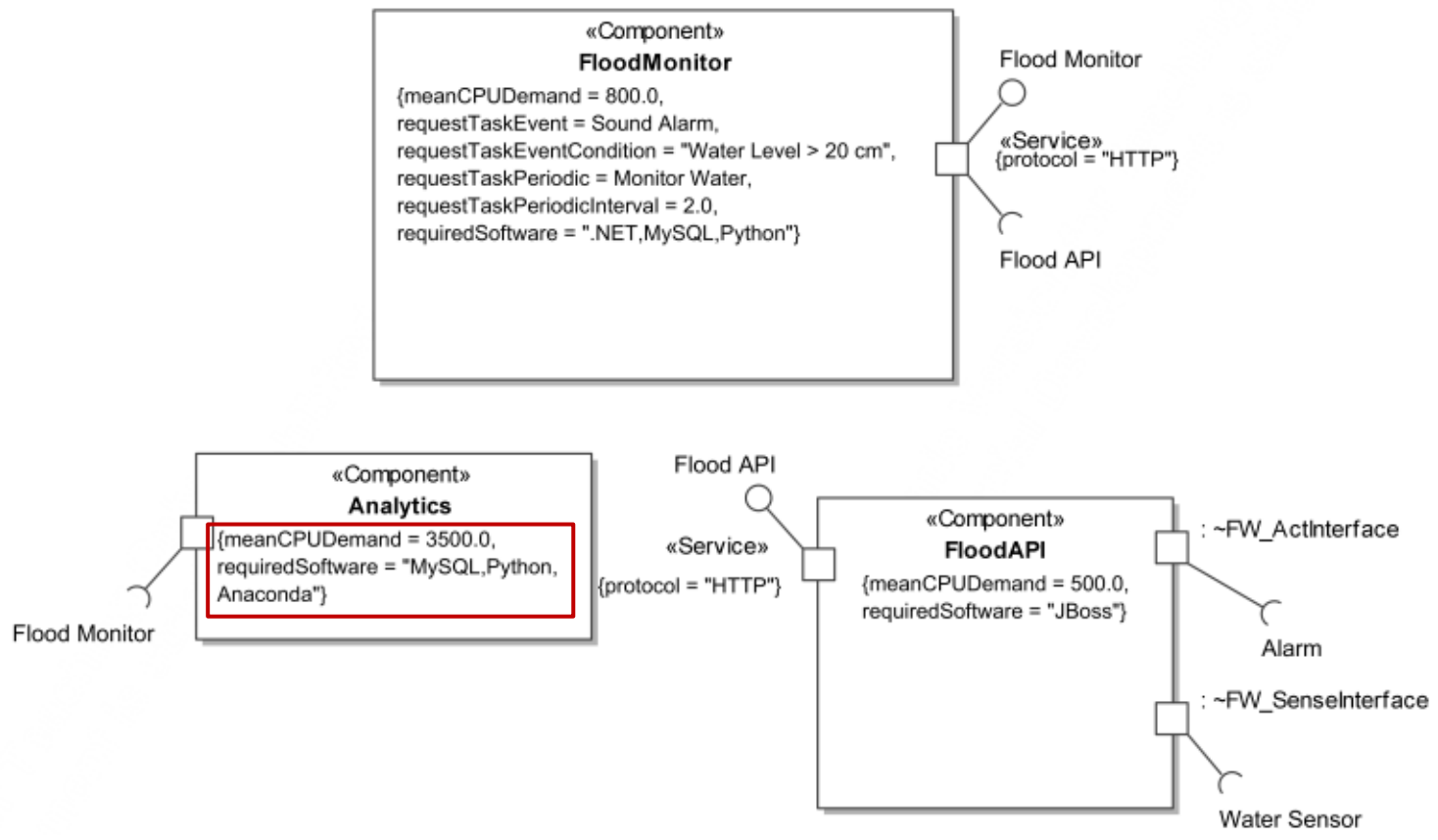

Figure 11. Components with "meanCPUdemand" and "requiredSoftware" information

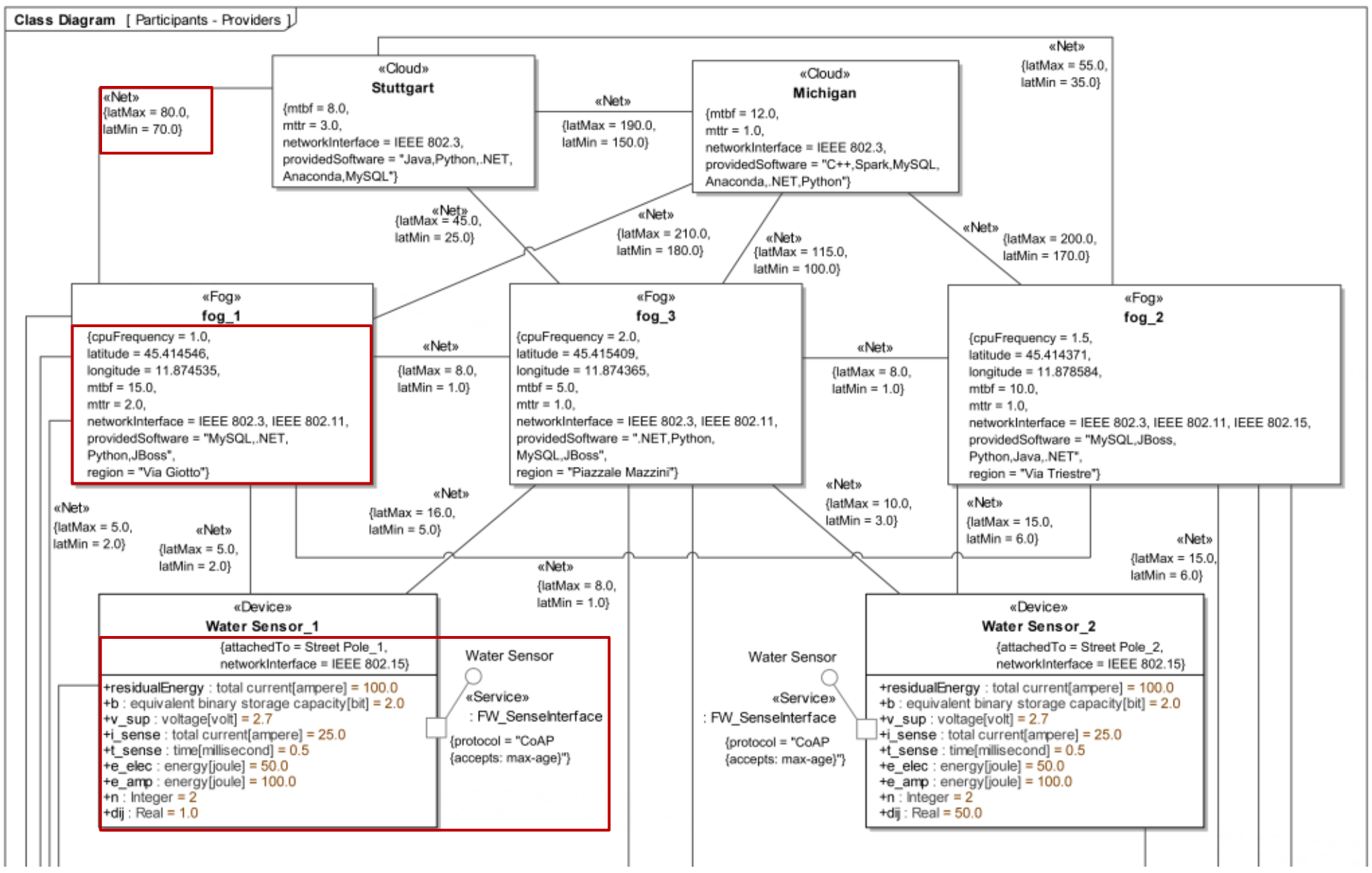

Figure 12. Providers and networks enhanced with information to analyze response time and availability

### 5.1.4.1 DQ1 and DQ2: Lifetime and Data freshness

Based on the participants design (Figure 12), water sensor devices can communicate with *fog_1*, *fog2*, or *fog_3*. Each of these fog nodes are located in a different place of the city of Padova. Thus, for each simulation scenario, we vary the distance of the devices, setting a random value from 1 to 50 (meters) for the attribute $d_{ij}$. After starting the model execution, we monitor the variable *residualEnergy*. When the value of the variable is above a given constant (i.e., 5 mAh), we stop the simulation and verify the value of the tagged value *globalTime*. We also vary the request rate from 2 to 6, aiming to analyze the lifetime of the water sensor device with different request rates. For each simulation scenario, we perform 30 execution rounds. The result is depicted in Figure 13 (a). The simulation shows that requesting the water level data every 2 seconds, the average lifetime of the devices is 9 months. On the other hand, by increasing the request rate to 4 and 6 seconds, the lifetime of the water sensor device increases to 12 and 15 months.

For analyzing the operational lifetime considering the data freshness required by the application, in the same way as the previous simulation scenario, we also vary the distance of the devices of the fog nodes. After starting the model execution, we monitor the variable *residualEnergy*. When the value of the variable is above a given constant (i.e., 5 mAh), we stop the simulation and verify the value of the tagged value *globalTime*. We also vary the max age of data, from 1 to 4 (seconds), aiming to analyze the lifetime of the water sensor device with different data freshness. For each simulation scenario, we perform 30 execution rounds. Figure 13 (b) shows the results. By allowing to request water level with higher max ages, the lifetime of the devices increases considerably. For example, in a scenario that the max age is 2 seconds, the lifetime of the devices increases in approximately 45% when compared with 1 second of max age. Whether the requests allow 4 seconds of max age, the operational lifetime could be more than duplicated.

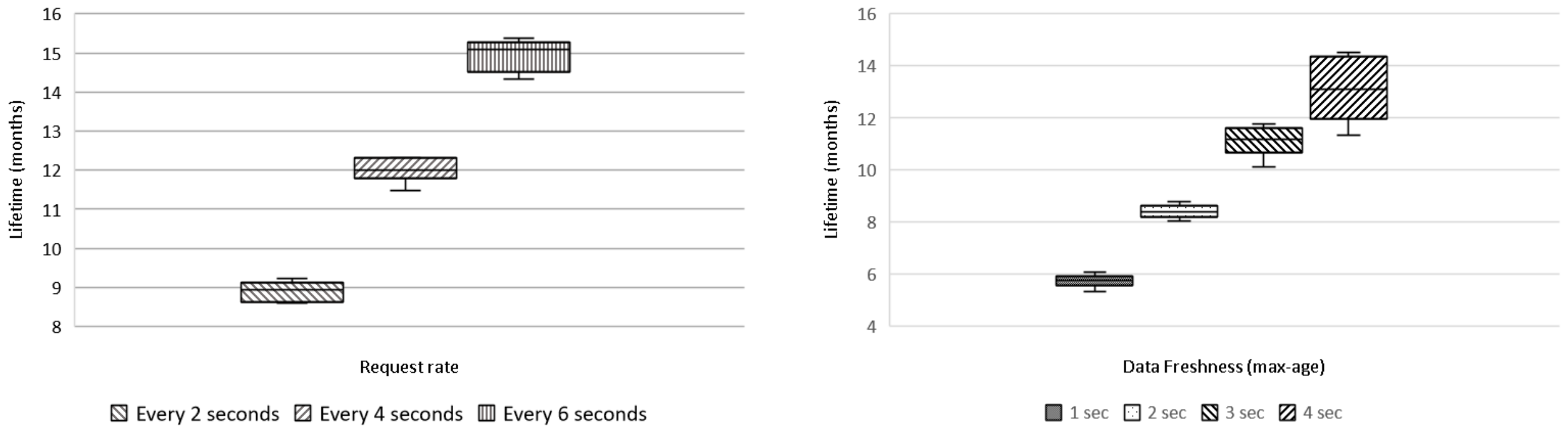

(a) (b)

Figure 13. Operational lifetime of water sensor devices considering different request rates of FW application (a), and Operational lifetime of water sensor devices considering different data freshness requirements (b)

#### 5.1.4.2 DQ3, DQ4, and DQ5: Deployment configurations, response time and availability.

By running the model, considering the software requirements, IoTDraw come up with 30 eligible deployments (recall that more than one component can be deployed in the same node). Part of the output is depicted in Figure 14. Each scenario is numbered, and, for each candidate, a deployment configuration is presented. For example, in *Scenario 1*, the *Analytics* and *FloodMonitor* components are deployed within *Stuttgart* while the *FloodAPI* component is deployed within fog_1. It means that such nodes provide all required software and hardware capabilities for the components.

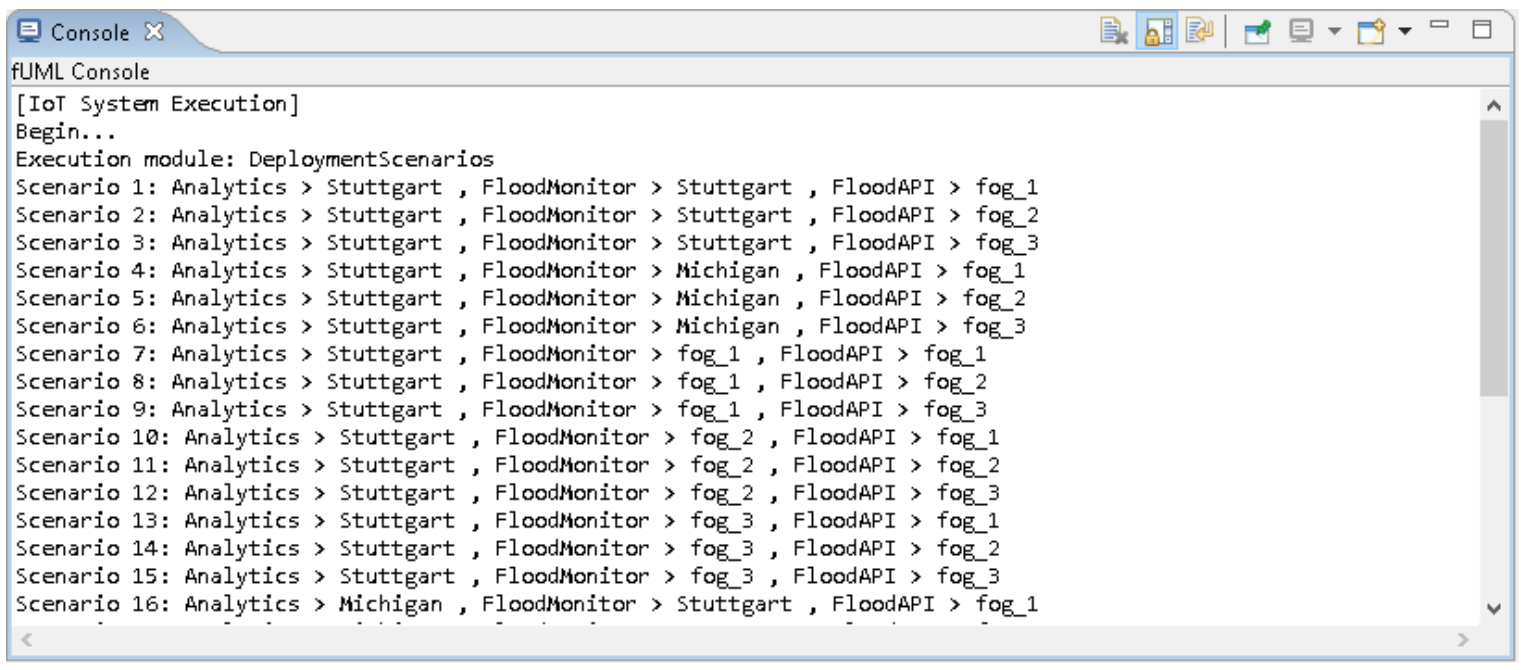

Figure 14. Eligible Deployment Candidate Scenarios

Despite the simulation helps to identify the possible deployment configurations, selecting a candidate scenario must consider two other requirements for the FW application, that is, availability and response time. Figure 14 depicts the deployment scenarios considering availability while Figure 15 depicts the eligible Deployment Candidate Scenarios ordered by response time.

The simulations help to reveal the best and worst deployment scenarios considering the availability and response time requirements. A conflict was identified, once the best scenario of response time (i.e., *Scenario 1*) is, oppositely, the third worst scenario when considering the availability. However, it was also possible to identify one scenario that meets both availability and response time, that is, *Scenario 19* (*Analytics*>*Michigan*, *FloodMonitor*>*Michigan*, *FloodAPI*>*fog_1*), which is the fifth best scenario of availability and the fourth best scenario for response time. Finding such a scenario without our supporting framework is a non-trivial task. For example, the average latency between the fog nodes and the *Stuttgart* cloud node is about 50 mS. On the other hand, the average latency between the fog nodes and the *Michigan* cloud node is about 160 mS. Thus, the likely decision would be deploying the components within *Stuttgart* cloud. However due to other features, e.g., software requirements and CPU speed, the best scenario addressing availability and response time both *Analytics* and *FloodMonitor* are deployed within *Michigan* cloud.

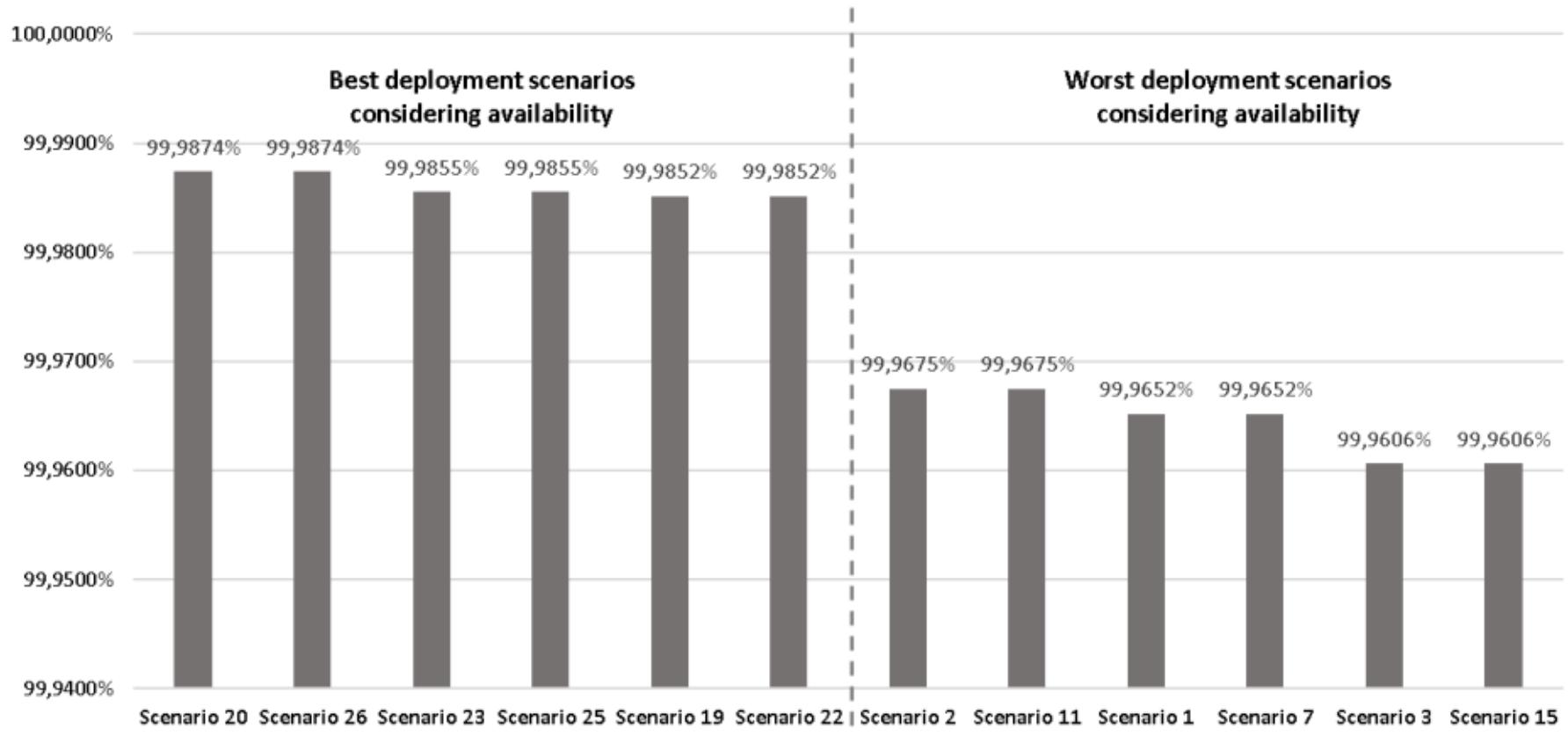

Figure 15 Deployment scenarios considering availability

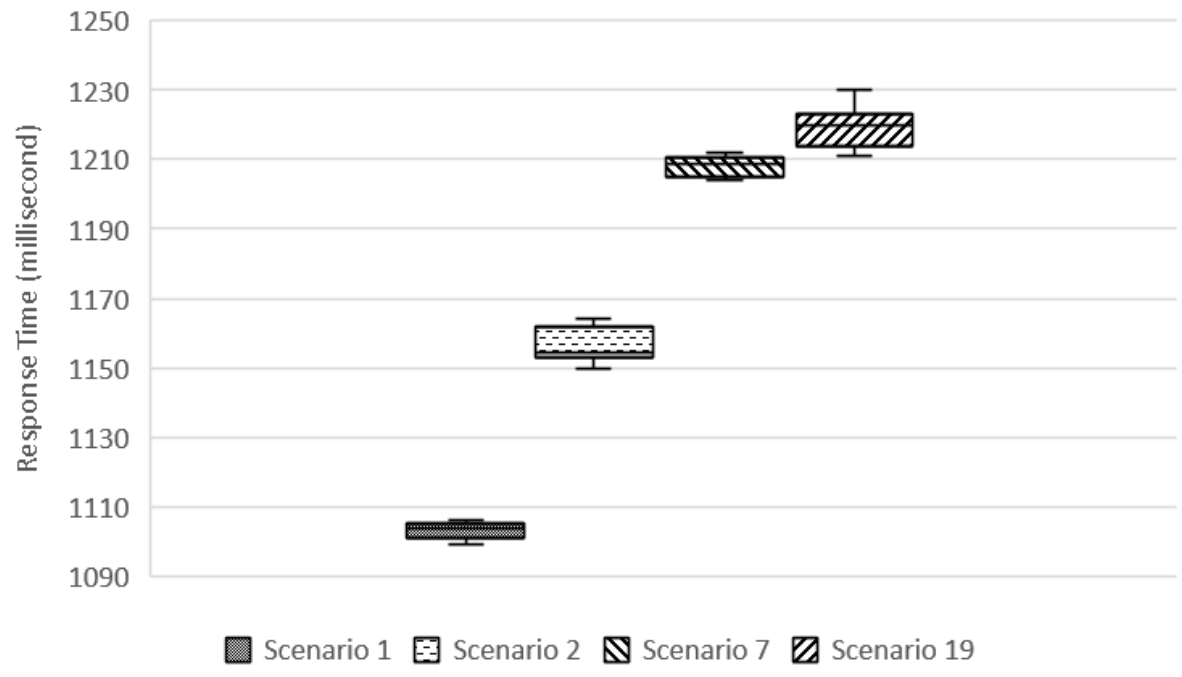

Figure 16. Deployment scenarios considering response time

### 5.1.5 Qualitative Analysis

A survey [24], conducted in the beginning of 2019, which reports on the interviews of more than 630 participants, from 250 companies spread around 42 countries, has identified some key concerns and barriers to develop successful IoT projects. The research revealed the following top five concerns: (i) the connectivity between the heterogeneous entities of the system; (ii) integration of hardware devices; (iii) interoperability between platforms; (iv) security, and; (v) total cost estimation. The IoTDraw aims to help the engineers tackle the connectivity, integration and interoperability concerns when developing IoT applications. In the following we discuss how the framework achieves this goal.

Connectivity is the backbone of IoT, which is achieved by the adoption of standards and protocols that establish how the software components and physical devices can connect and communicate between each other and with the Internet. Considering the connectivity between the heterogeneous entities composing an IoT system, SoaML4IoT stablishes that the modeler must define the protocols used in the communication paths and services. Thus, in the system architecture design, the services can only be connected whether their respective protocols are compatible. Such features of our proposed language aim at promoting the connectivity between system elements. For example, as depicted in Figure 7, the service of FloodMonitor component adopts the application protocol HTTP, while the protocol used in the service of Water Sensor_1 (Figure 8) was defined based on CoAP. In the system architecture (e.g., Figure 9), when creating the links between services, which are modeled trough communication links, both services (at each side of the link) must agree with the defined protocols.

When designing an IoT application in the IoTDraw, the modelers must specify the services' interfaces in a standardized fashion. This is achieved by the contract-based approach, which is adopted by SoaML4IoT. For example, in Section 5.1.1, we show how to specify contracts between service providers and consumers, and, in the architecture specification (Section 5.1.3) such contracts are used to verify whether the provider and consumer participants are compatible. Since the specification is realized through interfaces, it is agnostic of the type of entity that will implement it. Such approach is inspired by the SOA style and, as demonstrated by many studies (e.g., [67], [68]), promotes the integration and interoperability between the components of system (second and third concerns when developing IoT applications).

The current version of IoTDraw does not provide mechanisms to deal with security and cost estimation. However, such concerns point opportunities for enhancements in our proposed framework, which will be addressed in future works (Section 7).

### 5.2 Complex Scenario: New Sensors and Applications

After introducing and explaining IoTDraw by using a scenario with a limited number of components, in this section we introduce a more complex scenario. The purpose of this section is twofold. First, we aim at demonstrating the usefulness of IoTDraw to answer design questions that are humanly infeasible to be answered without a supporting tool. In this sense, as it will be described in the following, the scenario comes up with almost 1.000 eligible deployment configurations, each one impacting differently in the QoS requirements. Second, the scenario shows that our framework does not impose a strict division between cloud-fog-device levels for deployment. Actually, although we contemplate a 3-tier organization of IoT systems, as often adopted in the literature, the framework supports n-tier architectures. In order to illustrate such a feature, we modified the PSC architecture so that the components can be deployed in any cloud, fog or device nodes, considering the required and provided software, and the possible network connections.

We added several devices equipped with new sensors upon the PSC infrastructure. A small battery powers the devices and they are homogeneous, that is, equipped with only one type of sensor, which can be a photometer, a temperature sensor, a humidity sensor, or a benzene (C6H6) sensor (to monitor the air quality). The devices also have a CC2420 transceiver that implements the IEEE 802.15.4 standard [50], providing them with wireless communication capability. The new sensors are equipped with a more robust hardware, allowing them to provide execution environments for software components. Thus, the software components can be deployed within any processing node that makes part of the system. Figure 17 depicts the participant providers of the new scenario.

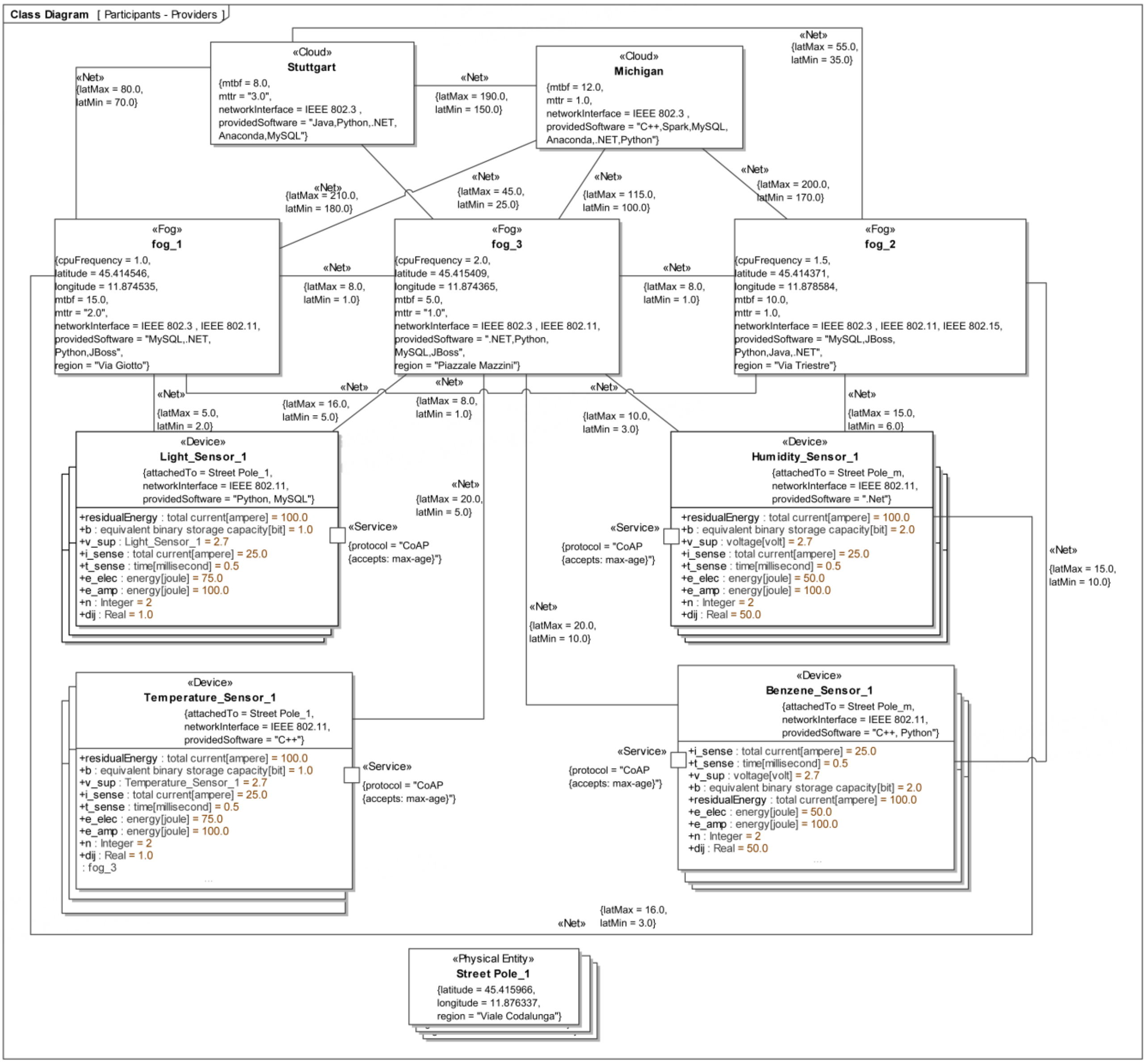

Figure 17. Participants - Providers of the new scenario

### 5.2.1 Application requirements

In this new scenario for running the PSC infrastructure, the devices endowed with different sensors enable the development of new applications aiming to enhance the quality of the services offered to citizens while reducing the operational costs of the public administration. The new applications we want to specify are the Street Light Monitoring (LM), the Temperature Monitoring (TM), the Humidity Monitoring (HM), and the Air Quality Monitoring (AM). The goal of such applications is to request the related data at regular time intervals (i.e., *periodic applications*), store it in a database, and further perform analytics aiming at examining different scenarios. Figure 18 depicts the requirements of each application specified by using the SoaML4IoT.

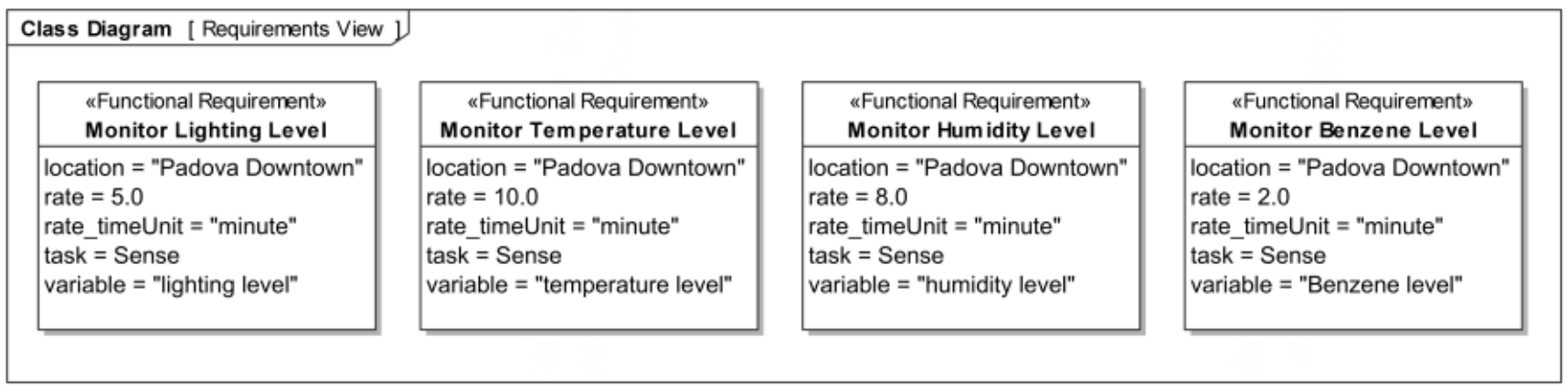

Figure 18 Requirements of the new applications of PSC

As with the flood warning (FW) application, constrained batteries power the devices providing the data for the new applications; thus, a QoS requirement that must be analyzed in this new scenario is the operational *lifetime* of devices providing the required services. Based on [55], we consider the lifetime as the time spanning from the instant when the device starts functioning until it runs out of energy, making the service provided by such device unavailable. In this sense, the ability of the service, platform, or component to perform its required function over an agreed period may be impacted. Thus, the LM, TM, HM, and AM applications shall also address *availability* requirements [15]. Although the new applications are not time-restricted, we want to conceive the applications' architecture providing the minimum delay in the communication between components. Based on [56], we consider the response time as the time spent since a component requests a service to another component until this first component receives the respective answer. We will reuse the specification of availability and response time conceived for the FW application, which is introduced in Section 5.1.4. Finally, the sensors will also be exposed as a service to be requested by external applications based on the *request-response* model, following the Sensing as a Service ($S^2$aaS) approach [54]. Thus, there is also a critical concern for the service consumers, namely, the *freshness* of the data. Data freshness refers to the time elapsed since the data is collected until it is delivered to the requesting user [57].

### 5.2.2 Participants Design

Since we have already introduced the diagrams modeling the service contracts (Section 5.1.2), in this scenario we focus on the specification of provider and consumer participants. Table 1 introduces the specification of participant consumers. To create each component, the modeler follows the same steps as presented in Section 5.1.2. For clarity, the new applications follow the same architecture of the FW application, that is, comprising three components: Monitor (managing the monitoring of the environmental variable), analytics (analysis module), and API (performing the requests to the services).

Table 1. Participant consumers

| Application | Component | meanCPUDemand | requiredSoftware |
|---|---|---|---|
| LM | MonitorLM | 800 CPU cycles | .NET, Python |
| | AnalyticsLM | 3.500 CPU cycles | Spark, C++ |
| | APILM | 500 CPU cycles | JBoss |
| TM | MonitorTM | 650 CPU cycles | .NET |
| | AnalyticsTM | 3.000 CPU cycles | Python, MySQL |
| | APITM | 550 CPU cycles | Python |
| HM | MonitorHM | 700 CPU cycles | Java |
| | AnalyticsHM | 2.500 CPU cycles | Anaconda |
| | APIHM | 490 CPU cycles | JBoss |
| AM | MonitorAM | 685 CPU cycles | Java |
| | AnalyticsAM | 2.530 CPU cycles | Spark |
| | APIAM | 250 CPU cycles | JBoss |

The participant providers refer to the new devices exposing the sensing data to the applications. All devices are placed on streetlight poles of Padova's downtown, following the same infrastructure introduced in Section 5.1. We consider 20 devices exposing the lighting level; 15 devices exposing the temperature level; 25 devices exposing the humidity level; and 5 devices exposing the benzene level. These former devices are placed only in streetlight poles that are near to the main boulevards of Padova.

Due to the wireless range, there are devices that can communicate with all fog nodes while other devices are connected only to a single fog node. To specify the participant providers, the modelers perform the same steps as introduced in Section 5.1. However, in this scenario, there are dozens of devices, which may hamper the specification of each one without the support of IoTDraw. Thus, our framework allows replicating the specification of an element, helping the modelers to create all elements composing the IoT system.

After specifying the consumers (i.e., applications) and providers (i.e., devices) participants, we want to answer design questions similar to the ones elicited for the FW application, that is, (**DQ1**) *considering the data sending rate required by the applications, what is the operational lifetime of the devices?*; (**DQ2**) *in the* $S^2$*aaS model, what is the operational lifetime of the devices considering the required data freshness of service consumers?*; (**DQ3**) *in which node, device cloud and/or fog, should the application's components be deployed?*, (**DQ4**) *what is the deployment configuration that promotes the highest availability for the applications?*, and; (**DQ5**) *what is the deployment configuration that promotes the lowest response time for the applications?* To answer these questions, we execute the model following the steps introduced in Section 5.1.4 in a computer with Intel Core i7-2640M, 2.80GHz, 6GB of RAM, JRE version 1.8.0_191.

Considering the first question, based on the participants' design, some of the devices can communicate only with *fog_1*, *fog_2*, or *fog_3*. On the other hand, other devices can communicate with two or even all fog nodes. Each of these fog nodes is located in a different place of the city of Padova. Thus, for each simulation scenario, we vary the distance of the devices, setting a random value from 1 to 50 (meters) for the attribute $d_{ij}$, which is the maximum wireless range of the devices. After starting the model execution, we monitor the variable *residualEnergy*. When the value of the variable is above a given constant (i.e., 5 mAh), we stop the simulation and verify the value of the tagged value *globalTime*. For each simulation scenario, we perform 30 execution rounds. The result is depicted in Figure 19 (a). Considering the data sending rates specified by the applications (Figure 18), the average lifetime of the devices providing data for the LM, TM, HM, and AM applications are 8, 21, 20, and 8 months, respectively.

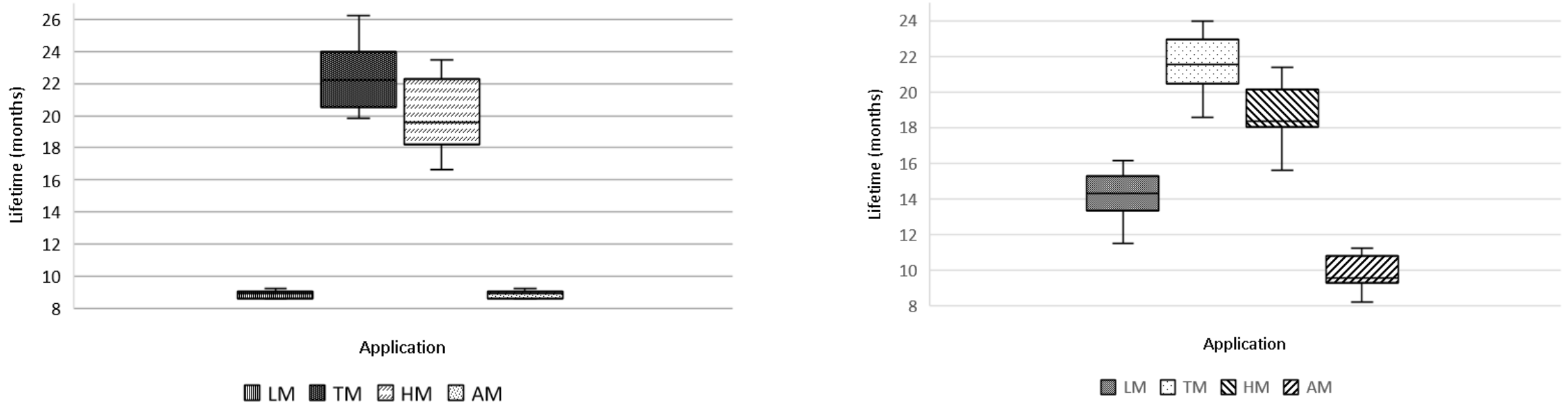

(a) (b)

Figure 19. (a) Lifetime of the applications LM, TM, HM and AM; (b) Lifetime of the applications LM, TM, HM and AM considering 4 seconds of data freshness

The second question aims at analyzing the lifetime of the devices considering the data freshness. In our analysis, we verify the lifetime by allowing up to 4 seconds of max-age of the data. The results are shown in Figure 19 (b). With the specified data rates (Figure 18), and allowing the data freshness up to 4 seconds, the average lifetime of the devices providing data for the LM, TM, HM, and AM applications are 14, 22, 20, and 10 months, respectively.

The objective of the third question is to find the eligible deployment scenarios considering network connections between the devices, provided and required software. Answering such question is important because each scenario may impact differently on the QoS requirements. To answer the question, we set the *DeploymentScenarios* execution module to the system model (Section 5.1.4) and execute it. Due to the number of scenarios, the execution spends about 3 minutes to come up with the results. Figure 20 depicts part of the result. IoTDraw comes up with 958 eligible deployment scenarios. Recall that each scenario considers: (i) the required and provided software and (ii) the possible network connections, that is, deployment scenarios consider only the ones that allow the communication between the components.

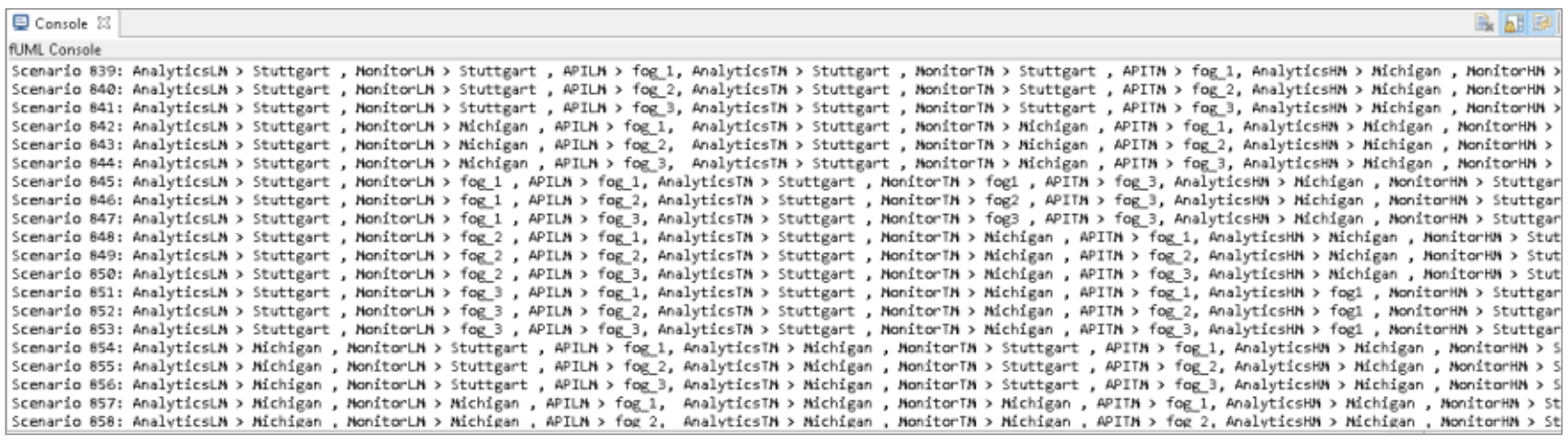

Figure 20. Deployment scenarios for the LM, TM, HM, and AM applications

With almost one thousand eligible deployment scenarios it becomes costly or even humanly infeasible to find the best deployment, since each option impacts differently on the availability QoS requirement. Finding the best option is crucial to guarantee the highest availability for the applications. This result is related to the fourth design question, that is, *what is the deployment configuration that promotes the highest availability for the applications?* To answer this question, we execute the module by setting the *AvailabilityAnalysis* to the system architecture. IoTDraw finds the best deployment configuration among the 958 eligible deployment scenarios, that is, the scenario 598:

```
Scenario 598:AnalyticsLM > Michigan, MonitorLM > fog_1, APILM > fog_3,
             AnalyticsTM > Stuttgart, MonitorTM > fog_3, APITM > fog_1,
             AnalyticsHM > Michigan, MonitorHM > fog_3, APIHM > fog_2,
             AnalyticsAM > Michigan, MonitorAM > fog_1, APIAM > fog_3
```

Considering the fifth question, recall that, the applications are composed of individual components which, in turn, can be deployed within different devices. Each device has a CPU frequency and the network connection between the devices have different latencies. At runtime, the components interact with each other to achieve the applications' goals. Thus, answering the fifth question allows us to find the deployment configuration that promotes the lowest response time in the components' requests. The execution was supported by the *ResponseTimeAnalysis* execution module, which was introduced in Section 5.1.4. Among hundreds of possible deployment configurations, IoTDraw finds the scenario that most reduces the response time between components' requests, that is, that scenario 168:

```
Scenario 168:AnalyticsLM > Michigan, MonitorLM > fog_2, APILM > fog_1,
             AnalyticsTM > Stuttgart, MonitorTM > Michigan, APITM > fog_3,
             AnalyticsHM > Michigan, MonitorHM > Stuttgart, APIHM > fog_2,
             AnalyticsAM > Stuttgart, MonitorAM > fog_2, APIAM > fog_2
```

### 5.3 Perceive of acceptance: usefulness and ease of use

The evaluation of perceive of acceptance is based on the Technology Acceptance Model (TAM) [62], a mature theoretical model that has been widely used in many empirical researches [69]. The TAM focuses on investigating the acceptance of a given system, which is determined by two criteria, that is, perceived usefulness (PU) and perceived ease of use (PEOU). The PU is defined as the degree to which a person believes that using the new technology will enhance their task performance [69]. On the other hand, PEOU refers to the degree to which a person believes that using a particular technology would be free from effort [62].

#### 5.3.1 Study Design

Our study is based on the evaluation model adopted by Espirito Santo [70], which, in light of [71], was designed as an adaptation of TAM by embodying the goal/question/metric paradigm (QGM) [72]. Figure 21 depicts the adapted model that we adopt in our study. The goals showed in the model are introduced in tables Table 2 and Table 3.

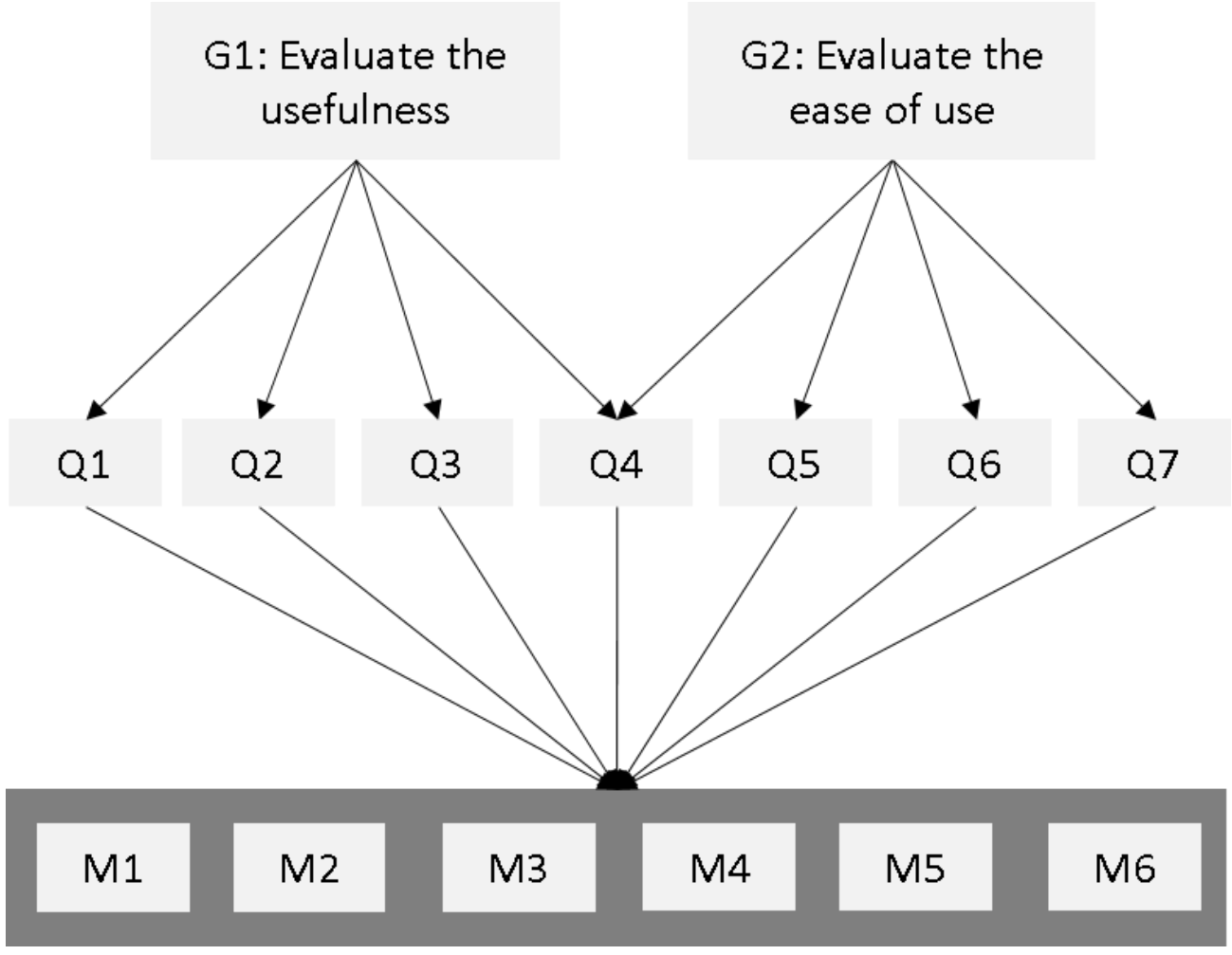

Figure 21.Usage of GQM in the TAM for evaluating the IoTDraw

Table 2. Goal G1

| | |
|---|---|
| **Analyze** | the IoTDraw framework |
| **Aiming to** | characterize |
| **Considering** | the usefulness of the tool |
| **In the context of** | the specification and analysis of SOA-based IoT applications |
| **Under viewpoint of** | software engineers modeling a SOA-based IoT application |

Table 3. Goal G2

| | |
|---|---|
| **Analyze** | the IoTDraw framework |
| **Aiming to** | characterize |
| **Considering** | the ease of use |
| **In the context of** | the specification and analysis of SOA-based IoT applications |
| **Under viewpoint of** | software engineers modeling a SOA-based IoT application |

Figure 21 depicts seven questions (Q1 to Q7) related to the objectives G1 and G2, following the GQM model. These questions, which are listed in Table 4, were conceived aiming at capturing the acceptance of IoTDraw in the dimensions of TAM, that is, its usefulness and ease of use. The possible answers were classified under an ordinal scale, that is, Totally Agree (TA), Broadly Agree (BA), Partially Agree (PA), Partially Disagree (PD), Broadly Disagree (BD), and Totally Disagree (TD).

Table 4. Questions for the evaluation of IoTDraw

| Question | Description | Dimension |
|---|---|---|
| **Q1** | It was easy to learn the IoTDraw framework | Ease of use |
| **Q2** | I was able to use the IoTDraw in the way I want to | Ease of use |
| **Q3** | I understand what happened in my interaction with IoTDraw | Ease of use |
| **Q4** | It was easy to specify and analyze the IoT application by using IoTDraw | Usefulness |
| **Q5** | I consider the IoTDraw framework useful to specify and analyze SOA-based IoT applications | Usefulness |
| **Q6** | The IoTDraw allows to specify and analyze an SOA-based IoT application following the activities defined in the development process | Usefulness |
| **Q7** | The usage of the IoTDraw allow enhancing the specification and analysis of SOA-based IoT applications | Usefulness |

Finally, to each question considered in the study, there is a set of metrics used to quantify them. Table 5 lists such metrics.

Table 5. Metrics for the evaluation of IoTDraw

| Metric | Description |
|---|---|
| **M1** | Number of participants that answer “Totally agree” |
| **M2** | Number of participants that answer “Broadly agree” |
| **M3** | Number of participants that answer “Partially agree” |
| **M4** | Number of participants that answer “Totally disagree” |
| **M5** | Number of participants that answer “Broadly disagree” |
| **M6** | Number of participants that answer “Partially disagree” |

To compute the final answer of each question of the study, it was considered the values related to each metric introduced earlier, following the configurations listed in Table 6.

Table 6. Configuration of the concept assigned to each question of GQM model

| Metric configuration | Assigned concept |
|---|---|
| **M1 > M2 + M3 ∨ M4, M5, M6 = 0** | Totally agree (TA) |
| **M1 ≤ M2 + M3 ∨ M2 > M3 ∨ M4, M5 = 0 ∨ M6 < M3** | Broadly agree (BA) |
| **M1 ≤ M2 + M3 ∨ M2 ≤ M3 ∨ M4 = 0 ∨ M6 < M3 ∨ M6 > M5** | Partially agree (PA) |
| **M1, M2, M3 = 0 ∨ M4 > M5 + M6** | Totally disagree (TD) |
| **M1, M2 = 0 ∨ M4 ≤ M5 + M6 ∨ M5 > M6 ∨ M6 > M3** | Broadly disagree (BD) |
| **M1 = 0 ∨ M4 ≤ M5 + M6 ∨ M5 ≤ M6 ∨ M6 > M3 ∨ M3 > M2** | Partially disagree (PD) |

Table 7. Interpretation model regarding the ease of use of IoTDraw

| Configuration | Interpretation |
|---|---|
| **Q1, Q2, Q3, and Q4 = TA** | IoTDraw is easy to use. It is not necessary improvements regarding usability, in such a way that the tool can be used immediately to support the specification and analysis of SOA-based IoT applications. |
| **Q1, Q2, Q3 and Q4 = BA, PA or PD** | IoTDraw is easy to use. However, the participants identified opportunities for improvements considering the ease of use. |
| **Q1, Q2, Q3 and Q4 = TD or BD** | IoTDraw is not easy to use. Therefore, the usability of the framework should be reviewed based on usability heuristics described in the technical literature and in the participants’ comments. |

Table 8. Interpretation model regarding the usefulness of IoTDraw

| Configuration | Interpretation |
|---|---|
| **Q4, Q5, Q6 and Q7 = TA** | IoTDraw is very useful for the specification and analysis of SOA-based IoT applications. It is not necessary to implement improvements regarding aspects of utility; thus, the framework can be used immediately to support the specification and analysis of SOA-based IoT applications. |
| **Q4, Q5, Q6 and Q7 = BA, PA or PD** | IoTDraw is very useful for the specification and analysis of SOA-based IoT applications. However, the participants identified opportunities for improvements considering the usefulness of the framework. |
| **Q4, Q5, Q6 and Q7 = TD or BD** | IoTDraw is not very useful for the specification and analysis of SOA-based IoT applications. In this way, the set of requirements that drove the development of the framework must be reviewed, aiming to identify the functionalities that may provide greater utility for the specification and analysis of SOA-based IoT applications. |

For the execution of the evaluation study of IoTDraw, we counted on the participation of 10 software engineers. All the participants have at least five years of experience in the specification and development of software systems in the industry. Furthermore, the participants have knowledge about UML, SOA, and the concepts of the Internet of Things, which has enabled them to develop prototypes of sensing applications with Arduino [73] and Raspberry Pi [74].

To support the evaluation, it was necessary to perform a training about basic functionalities of IoTDraw. Since the tool was conceived on Eclipse Papyrus, we also provide explanation and examples of usage of such a tool. In the training, we discuss IoT application domains based on the real applications developed by Libelium [36]. Quality attributes for IoT were also analyzed and discussed.

After the training, the participants were asked to conceive a hypothetical set of functional requirements for an SOA-based IoT application for a domain of choice. After this task, the participants were asked to specify an application by using the IoTDraw to fulfill such requirements. Next, we ask the participants to analyze the specified application in light of the current QoS attributes supported by the IoTDraw. Finally, the participants evaluate the framework by answering a Google Form. The answers were stored in a spreadsheet and analyzed according to the criteria introduced earlier.

### 5.3.2 Results

After performing the study and the participants answering the evaluation form, we analyze the answers. Firstly, we group them within the categories introduced before (i.e., TA, BA, PA, PD, BD, TD). Such grouping is presented in Table 9. Next, for each question, we apply the configuration model introduced in Table 7 and Table 8.

Table 9. Configuration of the concept assigned to each question of GQM model

| **Questions** | **Number of answers** | | | | | |
|---|---|---|---|---|---|---|
| | TA | BA | PA | PD | BD | TD |
| It was easy to learn the IoTDraw framework | 3 | 4 | 3 | 0 | 0 | 0 |
| I was able to use the IoTDraw in the way I want to | 3 | 2 | 4 | 1 | 0 | 0 |
| I understand what happened in my interaction with IoTDraw | 7 | 2 | 1 | 0 | 0 | 0 |
| It was easy to specify and analyze the IoT application by using IoTDraw | 2 | 5 | 2 | 1 | 0 | 0 |
| I consider the IoTDraw framework useful to specify and analyze SOA-based IoT applications | 8 | 1 | 1 | 0 | 0 | 0 |
| The IoTDraw allows to specify and analyze an SOA-based IoT application following the activities defined in the development process | 8 | 1 | 1 | 0 | 0 | 0 |
| The usage of the IoTDraw allow enhancing the specification and analysis of SOA-based IoT applications | 7 | 2 | 1 | 0 | 0 | 0 |

In general, it can be seen a trend of agreement with the questions proposed for evaluation of the IoTDraw. By classifying the answers according to the configurations presented in Table 7 and Table 8, we have the following results presented in Table 10. An important result of our study is that any participant broadly or totally disagrees with questions regarding ease of use and usefulness of IoTDraw. The improvements pointed by the participants will be addressed as future work.

Table 10. Interpretation model regarding the ease of use of IoTDraw

| Criteria | Interpretation | Agree |
|---|---|---|
| Ease of use | IoTDraw is easy to use. It is not necessary improvements regarding usability, in such a way that the tool can be used immediately to support the specification and analysis of SOA-based IoT applications. | **43.3%** |
| | IoTDraw is easy to use. However, the participants identified opportunities for improvements considering the ease of use. | **56.7%** |
| | IoTDraw is not easy to use. Therefore, the usability of the framework should be reviewed based on usability heuristics described in the technical literature and in the participants' comments. | **0%** |
| Usefulness | IoTDraw is very useful for the specification and analysis of SOA-based IoT applications. It is not necessary to implement improvements regarding aspects of utility; thus, the framework can be used immediately to support the specification and analysis of SOA-based IoT applications. | **62.5%** |
| | IoTDraw is very useful for the specification and analysis of SOA-based IoT applications. However, the participants identified opportunities for improvements considering the usefulness of the framework. | **37.5%** |
| | IoTDraw is not very useful for the specification and analysis of SOA-based IoT applications. In this way, the set of requirements that drove the development of the framework must be reviewed, aiming to identify the functionalities that may provide greater utility for the specification and analysis of SOA-based IoT applications. | **0%** |

## 6 Related Work

In this section, we analyze the related work in the context of IoT (UML4IoT [9], ArchWiSeN [10], COMFIT [6], Patel & Cassou [4], IoT Link [75], ThingML [20], and ELIoT [76]) and Cyber-Physical Systems (Zhang [77], TERRA [78], and Thramboulidis [79]), which can be seen as a superset of IoT. Initially, we introduce each study, highlighting their characteristics that are related to our work. Next, we analyze the existing works, compared to our proposal, in the light of two different aspects, which constitute key contributions of our work. The first aspect is the expressiveness of the SoaML4IoT metamodel in comparison with the other proposals. The second aspect regards how well our proposal and related studies fulfil a set of requirements that a modeling framework must meet in order to enable the precise representation and simulation of SOA-based IoT applications, as introduced by the literature.

The UML4IoT [9] consists of a UML profile comprising concepts for representing low-level mechatronic components and IoT sensors for manufacturing systems. The authors consider an IoT manufacturing system as a composition of cyber-physical, cyber and human components, and the IoT acts as glue for the integration of such components. The profile exploits the OMA LWM2M application protocol [80] and IPSO smart objects [81], which focus on modeling the exposed interface of simple smart objects. The DSL has implicit execution semantics. It provides a code generator of the IoT-compliant layer that is required for the cyber-physical component to be integrated into the analyzed IoT manufacturing environment.

The ArchWiSeN (Architecture for Wireless Sensor and Actuator Networks) [10] is an infrastructure based on Model-Driven Architecture (MDA) [8] for Wireless Sensor Networks. As a particular vision of MDD proposed by the OMG; thus, MDA is built on OMG standards (e.g., UML, MOF). It relies on the creation of models in different levels of abstraction (business, application, and platform technology), and on performing a set of *model-to-model* and *model-to-code* transformations. Thus, in MDA approaches, the execution semantics is essentially implicit in the translators. The authors of ArchWiSeN [10] propose a generic middleware metamodel as well as platform-specific metamodels aiming to support the code generation and UML-based DSLs.

Similarly to ArchWiSeN, the authors of COMFIT (Cloud and Model based IDE for the Internet of Things) [6] also propose a development environment built on MDA principles. It provides a graphical modeling language based on UML for designing applications focused on WSN applications. The framework encompasses a generic middleware metamodel including concepts shared among different WSN solutions. The objective is to generate middleware models tailored to the application requirements while respecting the resource constraints of the devices.

Patel and Cassou [4] present an MDD framework for IoT applications that addresses the lack of division of roles and the heterogeneity of devices in different phases of the IoT application development life cycle. The authors identify the stakeholders in WSN systems and propose a set of textual DSLs providing abstractions for specifying different types of devices of the WSN domain. The execution semantics of these languages are defined implicitly, through code-generators that translate the models into the formal target languages.

The authors of IoTLink [75] propose a framework to help inexperienced developers to create IoT applications. To achieve this goal, IoTLink provides graphical DSLs based on the MDA that encapsulate the complexity of communicating with devices and services on the Internet. Furthermore, the proposed approach also abstracts devices as virtual objects that are accessible through different communication technologies. A code generator specifies the execution semantics, by translating the compliant models into Java code.

The ThingML [20] framework consists of a textual modeling language and a set of tools for developing IoT applications. The modeling language is aligned with UML by applying concepts of statecharts and components. The tools encompass code generators to translate the models created with the proposed language into formal target languages. These tools comprise the implicit execution semantics of ThingML.

ELIoT [76] is an IoT application development platform that provides a textual DSL comprising concepts to represent Internet-connected smart devices. Unlike the other approaches, the execution semantics of the DSL provided by ELIoT is defined explicitly. Thus, as allowed by IoTDraw, system models created with such a language can be executed in a run-time engine platform provided by the approach. The DSL provided by ELIoT is based on Erlang [82], a programming language used to build real-time systems. Thus, specifying IoT applications with ELIoT assumes that the user has intermediate knowledge of Erlang.

In the context of CPS, in [77] the author proposes an approach to support the specification and modeling of automotive cyber-physical systems based on systems-of-systems engineering. A formal specification method was applied in the requirement analysis process in order to ensure that the software requirements model satisfies the required system function, performance goals and constraints, including safety. The author applied ModelicaML, whose execution semantics is defined explicitly, allowing the execution of ModelicaML-based models. The metamodel proposed by the authors encompasses low-level, electronic components of IoT devices.

Similar to the previous work, the Twente Embedded Real-time Robotic Application (TERRA) [78] is an MDD tool suite, supporting the design of IoT applications that also focuses on electronic components of IoT devices. Its tools range from a DSL and editors to graphically design the models to code generation tools to convert the models into source code. All structures of the TERRA models are defined by meta-models. These meta-models define all model elements and their usage. TERRA adopts code generation to execute the models, thus, the execution semantics is defined implicitly in the translator.

Finally, the work proposed by Thramboulidis [79] introduces a modeling framework for Industrial automation systems (IASs). The software for IAS commonly integrates mechanics and electronics. The framework is based on a cyber-physical system-based approach and can also be used to generate implementations on the various ARM-based embedded boards that have recently appeared in the market. These implementations are performed through a translation from the models into the target boards, thus, applying the implicit semantics to simulate the models.

### 6.1 Expressiveness analysis

The expressiveness concerns the degree to which a language allows its users to capture phenomena in the domain [83]. Thus, the more aspects of the domain of interest can be represented by the language, the higher expressiveness such language has. In light of the framework for measuring expressiveness in conceptual modeling proposed by Patig [84], we assess the expressiveness of IoTDraw metamodel comparing it with the related work introduced above.

In the Patig's framework, $D_1$ and $D_2$ denote two metamodels, whose expressiveness is to be compared. The models are represented by $M(D_1)$ and $M(D_2)$, respectively. Two metamodels $D_1$ and $D_2$ are *equally expressive* if every model of $D_1$ is also a model of $D_2$ and vice-versa, i.e., $M(D_1) = M(D_2)$. On the other hand, a description $D_1$ is *more expressive* than a description $D_2$ if all models of $D_2$ are proper subsets of the models of $D_1$, that is, $M(D_1) \supset M(D_2)$. In other words, $D_1$ provides all elements to create $M(D_2)$, including the elements to create $M(D_1)$. In our evaluation, we compare the expressiveness of $M(\text{IoTDraw})$, $M(\text{UML4IoT})$ [9], $M(\text{ArchWiseN})$ [10], $M(\text{COMFIT})$ [6], $M(\text{Patel\&Cassou})$ [4], $M(\text{IoT Link})$ [75], $M(\text{ThingML})$ [20], $M(\text{ELIoT})$ [76], $M(\text{Zhang})$ [77], $M(\text{TERRA})$ [78], and $M(\text{Thramboulidis})$ [79].

The expressiveness of the metamodels are verified according to a set of statements that the models can represent. By statement, we mean a syntactic expression and its meaning. As introduced by the Patig's framework, first-order predicate logic [85] can be used as a common description language to represent the statements, because of the comprehensible definition of its models. Thus, each statement is modeled as an atomic formula, representing one a portion description of the domain of interest.

In order to conceive the list of statements to compare the expressiveness of SoaML4IoT metamodel with related works, we analyze proven reference models of IoT and SOA. Reference models represent a set of essential concepts and relationships between them concerning a specific domain abstractly [86]. They are based on a small number of unifying concepts and can be used as a basis for explaining a given domain. We identified key concepts covered by such reference models and represent them as first-order logic statements. For the IoT domain, we analyze the IoT Architectural Reference Model (IoT-ARM) [11], the WSO2 IoT Reference Architecture [28], and the IEEE Standard for an Architectural Framework for Internet of Things (P2413) [58]. Considering the SOA style, we analyze the OASIS [87] and SOA Ontology [88], which are two of the most popular reference models for SOA.

The set of statements is listed in Table 11. Besides identifying key elements of the IoT domain, these statements represent the expressivity of the metamodel to allow answering design questions related to the SOA style.

Table 11. Statements for SOA-based modeling frameworks

| # | Statement | Description |
|---|---|---|
| ***S1*** | $iotSystem(i)$ | The IoT system $i$ |
| ***S2*** | $platform(p)$ | The platform $p$ |
| ***S3*** | $cloud(p)$ | The platform $p$ is of type cloud |
| ***S4*** | $fog(p)$ | The platform $p$ is of type fog |
| ***S5*** | $device(d)$ | The platform $p$ is of type device |
| ***S6*** | $physicalEntity(e)$ | The physical entity $e$ |
| ***S7*** | $location(l, e)$ | The location $l$ of physical entity $e$ |
| ***S8*** | $isAttachedTo(d, e)$ | The device $d$ is attached to the physical entity $e$ |
| ***S9*** | $task(t, d)$ | The task $t$ performed by the device $d$ |
| ***S10*** | $network(n)$ | The network $n$ |
| ***S11*** | $application(a)$ | The application $a$ |
| ***S12*** | $component(c, a)$ | The component $c$ of the application $a$ |
| ***S13*** | $consumer(c)$ | The role $m$ of the component $c$ |
| ***S14*** | $provider(d)$ | The role of the device $d$ |
| ***S15*** | $service(t, s)$ | The task $t$ is exposed by the service $s$ |
| ***S16*** | $contract(s, t)$ | The contract $t$ of the service $s$ |
| ***S17*** | $choreography(s, h)$ | The choreography $h$ of the service $s$ |
| ***S18*** | $messageType(s, m)$ | The message type $m$ of the service $s$ |
| ***S19*** | $requirement(r)$ | The requirement $r$ |
| ***S20*** | $functional(r, f)$ | The type $f$ (Functional) of requirement $r$ |
| ***S21*** | $qos(r, q)$ | The type $q$ (QoS) of requirement $r$ |
| ***S22*** | $mustFulfill(a, r)$ | The application $a$ must fulfill the requirement $r$ |
| ***S23*** | $simulation(u, i)$ | The simulation $u$ of an IoT system $i$ |
| ***S24*** | $timer(t, u)$ | The simulation timer $t$ of the simulation $u$ |
| ***S25*** | $event(v)$ | The simulation event $v$ |
| ***S26*** | $stateVariable(x)$ | The state variable $x$ |
| ***S27*** | $state(y)$ | The simulation state $y$ |
| ***S28*** | $activity(z)$ | The simulation activity $z$ |

After introducing the statements, in Table 12 we compare the expressivity of our proposal with the existing modeling frameworks for IoT, with focus on their capability to represent SOA-based concepts. In our analysis, we search for concepts that are first-class elements in the modeling languages.

Table 12. Expressivity comparison between our approach and related work

| # | $\boldsymbol{M}$(**IoTDraw**) | **[9]** | **[10]** | **[6]** | **[4]** | [75] | **[20]** | [76] | [77] | [78] | [79] |
|---|---|---|---|---|---|---|---|---|---|---|---|
| ***S1*** | ✓ | ✓ | ✓ | ✓ | ✓ | ✓ | ✓ | ✓ | ✓ | ✓ | ✓ |
| ***S2*** | ✓ | ✓ | ✓ | ✓ | ✓ | ✓ | ✓ | ✓ | ✓ | ✓ | ✓ |
| ***S3*** | ✓ | - | - | ✓ | - | ✓ | - | - | - | - | - |
| ***S4*** | ✓ | ✓ | ✓ | ✓ | ✓ | ✓ | ✓ | ✓ | - | - | - |
| ***S5*** | ✓ | ✓ | ✓ | ✓ | ✓ | ✓ | ✓ | ✓ | ✓ | ✓ | ✓ |
| ***S6*** | ✓ | ✓ | - | - | - | ✓ | - | - | - | ✓ | ✓ |
| ***S7*** | ✓ | ✓ | ✓ | ✓ | ✓ | ✓ | ✓ | ✓ | ✓ | ✓ | ✓ |
| ***S8*** | ✓ | - | - | - | - | - | ✓ | ✓ | - | ✓ | - |
| ***S9*** | ✓ | ✓ | ✓ | ✓ | ✓ | ✓ | ✓ | ✓ | ✓ | ✓ | ✓ |
| ***S10*** | ✓ | ✓ | ✓ | ✓ | ✓ | ✓ | ✓ | ✓ | ✓ | ✓ | ✓ |
| ***S11*** | ✓ | ✓ | ✓ | ✓ | ✓ | ✓ | ✓ | ✓ | ✓ | ✓ | ✓ |
| ***S12*** | ✓ | - | - | - | - | - | ✓ | - | ✓ | ✓ | ✓ |
| ***S13*** | ✓ | - | - | - | - | - | - | - | - | - | - |
| ***S14*** | ✓ | - | - | - | - | - | ✓ | ✓ | ✓ | - | - |
| ***S15*** | ✓ | - | - | - | - | - | ✓ | ✓ | - | - | - |
| ***S16*** | ✓ | - | - | - | - | - | - | - | - | - | ✓ |
| ***S17*** | ✓ | - | - | - | - | - | - | - | - | - | - |

| # | $M$(IoTDraw) | [9] | [10] | [6] | [4] | [75] | [20] | [76] | [77] | [78] | [79] |
|---|---|---|---|---|---|---|---|---|---|---|---|
| ***S18*** | ✓ | - | - | - | - | - | ✓ | ✓ | - | ✓ | - |
| ***S19*** | ✓ | - | ✓ | ✓ | - | - | - | - | - | - | - |
| ***S20*** | ✓ | - | ✓ | ✓ | - | - | - | - | - | - | - |
| ***S21*** | ✓ | - | ✓ | ✓ | - | - | - | - | - | - | - |
| ***S22*** | ✓ | - | ✓ | ✓ | - | - | - | - | - | - | - |
| ***S23*** | ✓ | - | - | - | - | - | - | ✓ | - | - | - |
| ***S24*** | ✓ | - | - | - | - | - | - | ✓ | - | - | - |
| ***S25*** | ✓ | - | - | - | - | - | - | ✓ | - | - | - |
| ***S26*** | ✓ | - | - | - | - | - | - | ✓ | - | - | - |
| ***S27*** | ✓ | - | - | - | - | - | - | ✓ | - | - | - |
| ***S28*** | ✓ | - | - | - | - | - | - | ✓ | - | - | - |

✓= the framework provides concepts to represent the statement -= the framework does not provide concepts to represent the statement.

As presented in Table 12, it can be observed that SoaML4IoT is more expressive than the related works concerning the analyzed aspects. Most of the modeling frameworks provide concepts to represent elements of the IoT domain, that is, platforms (cloud, fog, device), tasks, networks, applications, and physical entities. However, many of the frameworks organize their applications as a monolithic software unit, i.e., they do not allow representing the application as a composition of components. Considering the concepts regarding SOA, none of the analyzed approaches provides first-class elements to represent participants roles, contracts, or choreography. The framework ThingML [20], indeed, is based on the services approach, however, it does not support a clear definition of aspects related to service contract and choreography.

Considering aspects related to requirements specification, only the approaches ArchWiseN [10] and COMFIT [6] provide components for representing functional and non-functional requirements. Such frameworks are based on MDA [8], in which the Platform-Independent Model (PIM) receives as an input the requirements that will define the behavior of the system. In the following, the PIM is transformed into the Platform-Specific Model (PSM), which specifies the technological aspects in detail aiming to address the requirements elicited earlier. The other approaches assume that application logic and application requirements have been previously defined and does not provide components related to functional and non-functional requirements.

Finally, only the framework ELIoT [76] provides concepts related to model execution and simulation, which allows simulating its models without requiring transforming its compliant models into another language. As a simulator, the framework allows representing the concepts of computational simulation, such as, timer, event, state and activity.

It can therefore be concluded that our proposal allows representing a wide range of characteristics of both the IoT and SOA domains, making it the most expressive approach for building SOA-based IoT applications, to the best of our knowledge.

### 6.2 Modeling framework analysis considering requirements elicited by the literature

Besides expressiveness, the literature indicates other elements that a modeling framework must provide in order to allow the precise representation and simulation of systems in general and, more specifically, SOA-based IoT applications. Such elements can be seen as requirements that a modeling framework must fulfill. The requirements are introduced by Brambilla and others [8] and complemented by Tatibouët and others [43]. In the following, we introduce each requirement, followed by the analysis of their fulfillment by IoTDraw and the related work.

**Expressiveness (MFR1)**: The first requirement that shall be addressed by an MDD framework is to provide a metamodel with a coherent set of concepts related to the IoT domain in general and, more specifically, the SOA style. The concepts shall be represented as first-class elements, supporting the identification of key components of an SOA-based IoT application. The concepts from both IoT and SOA domain must have an explicit relation, allowing to identify the roles of the IoT components considering the services approach.

**Well-Defined Notation (MFR2)**: Aiming to allow the representation of IoT systems, the second requirement that shall be addressed by a modeling framework is to provide a well-defined notation. It must cover the elements of its metamodel and provide a tool to support the specification of IoT systems by using the DSL.

**Extensibility (MFR3)**: The third requirement that shall be addressed by a modeling framework is a clear definition on how to extend it to allow representing other aspects that are not covered by the DSL.

**Explicit Execution Semantics (MFR4):** To execute and simulate the model allowing answering questions at design-time, the DSL must provide the execution semantics explicitly, avoiding applying translational approaches, and supporting the architectural decision-making process. Therefore, the third requirement for a modeling framework is that the execution semantics of the modeling language must be explicitly specified.

Table 13 provides an overview of the fulfillment of the modeling frameworks regarding the requirements MFR1 to MFR4, that is, Expressiveness (MFR1), Well-Defined Notation (MFR2), Extensibility (MFR3), and Explicit Execution Semantics (MFR4). From the table, we can first observe that none of the frameworks entirely fulfills our requirements.

Table 13. Requirements fulfilling of related work

| **Framework** | **MFR1** | **MFR2** | **MFR3** | **MFR4** |
|---|---|---|---|---|
| IoTDraw | ✓ | ✓[4,5] | ✓ | ✓ |
| UML4IoT [9] | o[2] | ✓[4,5] | o [6] | ✗[7] |
| ArchWiSeN [10] | o[2] | ✓[4,5] | o [6] | ✗[7] |
| COMFIT [6] | o[2] | ✓[4,5] | o [6] | ✗[7] |
| Patel & Cassou [4] | o[1] | ✓[3] | ✗ | ✗[7] |
| IoTLink [75] | o[1] | ✓[4,5] | o [6] | ✗[7] |
| ThingML [20] | o[1] | ✓[3,5] | o [6] | ✗[7] |
| ELIoT [76] | o[1] | ✓[3] | ✗ | ✓ |
| Zhang [77] | o[1] | ✓[4,5] | o [6] | ✗[7] |
| TERRA [78] | o[1] | ✓[3] | ✗ | ✗[7] |
| Thramboulidis [79] | o[1] | ✓[4,5] | o [6] | ✗[7] |

✓= requirement fulfilled; ✗= requirement not fulfilled o = requirement partially fulfilled; [1] provides concepts of IoT but lack SOA-related elements; [2] focus on WSN; [3] textual; [4] graphical; [5] based on UML; [6] it is possible to extend due the subjacent technology used to create the DSL, however, the approach has unclear extensibility rules; [7] implicit execution semantics.

We can observe that IoTDraw fulfills all the requirements for modeling frameworks. In the previous section, we analyze the expressivity of IoTDraw, when comparing it with other approaches. As introduced in Section 4.2, our framework provides a DSL based on UML, the SoaML4IoT. Section 4.4 introduces the extensibility rules of SoaML4IoT. Finally, as presented in Section 4.3, SoaML4IoT provides an explicit execution semantics, which allows the execution of SoaML4IoT-based models. The analysis of how the frameworks proposed by related works fulfill the abovementioned requirements is presented below.

The "Expressiveness" requirement, MFR1, is partially fulfilled by all the surveyed modeling frameworks. However, the approaches provide conceptualizations for either IoT systems in general, without concepts related to the SOA style, or specifically for Wireless Sensor Networks (WSN). When providing only general concepts for IoT, the approaches lack concepts to identify elements of SOA-based systems. On the other hand, modeling approaches that focus on WSN have a comprehensive set of concepts and related properties regarding devices and networks. In this sense, it is possible to represent several aspects of the infrastructure of the IoT system. However, without elements regarding upper abstraction levels, such as the applications' architecture, the model can be incomplete, lacking the support for the representation and analysis of the system. The requirement MRF2 "Well-Defined Notation" is also fulfilled by all the surveyed modeling frameworks. Most of them provide graphical notation representing the domain concepts. An important finding is that most of the graphical notations is based on UML, which reinforced our decision to use OMG standards to build our work on.

Regarding the third requirement (MFR3) – "Extensibility," an interesting finding is that none of the surveyed approaches provides clear specifications of extensibility rules to extend the DSLs, which would allow representing other aspects that the language does not provide. The extension of DSLs built on UML can follow the profiling mechanism. Without formalizing extensibility rules of the approaches, the authors transfer the extension process of their languages to the modelers, who may extend the language in an ad-hoc fashion. However, this is risky, since the extension may cause the DSL to become incompatible with the metamodel and other components of the modeling frameworks.

Finally, regarding the fourth requirement (MFR4) – "Explicit Execution Semantics" – we identify that only ELIoT [76] provides explicit execution semantics. On the other hand, all the other surveyed approaches provide DSLs with implicit execution semantics; thus, requiring translating the system description into a formal language to execute it. Another interesting finding is that none of the approaches focuses on supporting the architectural decision-making process. That is, the approaches do not provide mechanisms to predict at design time the properties of systems' artefacts. Instead, the proposals focus on generating code for the platforms. However, in this way, it is difficult to answer design questions and analyze the impact of design decisions on the architecture, since the execution models refer to the application code itself. To execute and simulate the system behavior it would be necessary to deploy the components in the platforms and verify the various design alternatives at runtime.

## 7 Final Remarks and Future Work

Through IoTDraw, the modelers would be able to identify key elements regarding both IoT and SOA style, considering the specification of periodic applications. Our framework helps to answer important design questions, supporting the architectural decision-making process. Even for the simple scenario of FW application, with a reduced number of platforms and components, it would be difficult to answer the design questions without the support of the IoTDraw. Considering a more complex scenario, it may become humanly infeasible to select the best architectural configuration without a supporting framework. Finally, the study conducted with software engineers of the industry has identified that our framework is easy to be used and, primarily, useful for the specification and analysis of SOA-based periodic IoT applications.

The SoaML4IoT extends the SoaML profile by enhancing its metamodel with IoT-specific concepts. Therefore, we follow the SoaML design, which does not specify any constraint to support model validation. In our approach, the only constraints that support the modelers to conceive models that are compliant with SoaML4IoT metamodels are the relationships and cardinalities. Important future work is to analyze which constraints should be created furthering the correctness of the models. Such constraints must reflect rules of the IoT applications domain. Since SoaML4IoT is fully OMG-compliant, the constraints can be written in OCL, for example, as invariants and pre- and post-conditions. Another direction is to advance the framework to support security and cost estimation specification and analysis, as such aspects were pointed as important concerns when developing IoT applications. Supporting security issues will require a deeper analysis, while concerning the cost estimation feature, we envision the possibility of extending SoaML4IoT by adding new tagged values representing the cost model of the provider. The calculation could be performed by creating a fUML Activity Diagram, and integrate it with the application behavior (e.g., Figure 6).

The current version of IoTDraw does not cover elements related to self-configuration. In IoT systems that encompass dozens to thousands of interacting IoT devices, this is an important ability to allow the system to adapt itself to environmental changes. Therefore, extend or integrate SoaML4IoT with reference models that provide concepts related to self-configuration, such as MAPE-K [89], is one important direction of this work. Finally, we envision the integration of SoaML4IoT with microservices metamodels. Thus, it would be possible to specify architecture with vertical decomposition, allowing to define an arbitrary number of hierarchical levels and a more fine-grained deployment optimization.

### Acknowledgments

This study was partially supported by the Coordenação de Aperfeiçoamento de Pessoal de Nível Superior – Brasil (CAPES), through Finance Code 001, and by São Paulo Research Foundation – FAPESP, through grant number 2015/24144-7. Paulo Pires and Flávia Delicato are CNPq Fellows